\documentclass[9.5pt,journal,compsoc]{IEEEtran}
\ifCLASSOPTIONcompsoc
\else
\fi
\usepackage{subfigure}
\ifCLASSINFOpdf
   \usepackage[pdftex]{graphicx}
   \graphicspath{{./figure/}}
   \DeclareGraphicsExtensions{.pdf,.jpeg,.png}
\else
\fi
\begin{document}
%

\title{HMTT: A Hybrid Hardware/Software Tracing System for Bridging Memory Trace's \\Semantic Gap}

%
%
%
%

\author{Yungang~Bao, Jinyong~Zhang, Yan~Zhu, Dan~Tang, \\
       Yuan~Ruan, Mingyu~Chen, and~Jianping~Fan\\
Institute of Computing Technology, Chinese Academy of Sciences\\
\{baoyg, cmy, tangdang, fan\}@ict.ac.cn~~\{zhangjinyong, zhuyan, ry\}@ncic.ac.cn
}

%
%

\markboth{CAS-ICT-Tech-Report-20090327}%
{Shell \MakeLowercase{\textit{et al.}}: HMTT: A Hybrid
Hardware/Software Tracing System for Bridging Memory Trace's
Semantic Gap}
%



\IEEEcompsoctitleabstractindextext{%
\begin{abstract}
Memory trace analysis is an important technology for architecture
research, system software (i.e., OS, compiler) optimization, and
application performance improvements. Hardware-snooping is an
effective and efficient approach to monitor and collect memory
traces. Compared with software-based approaches, memory traces
collected by hardware-based approaches are usually lack of semantic
information, such as process/function/loop identifiers, virtual
address and I/O access. In this paper we propose a hybrid
hardware/software mechanism which is able to collect memory
reference trace as well as semantic information. Based on this
mechanism, we designed and implemented a prototype system called
HMTT (Hybrid Memory Trace Tool) which adopts a DIMM-snooping
mechanism to snoop on memory bus and a software-controlled tracing
mechanism to inject semantic information into normal memory trace.
To the best of our knowledge, the HMTT system is the first hardware
tracing system capable of correlating memory trace with high-level
events. Comprehensive validations and evaluations show that the HMTT
system has both hardware's (e.g., no distortion or pollution) and
software's advantages (e.g., flexibility and more information).
\end{abstract}


}

\maketitle

\IEEEdisplaynotcompsoctitleabstractindextext

%
\IEEEpeerreviewmaketitle

\section{Introduction}
%
%

%
%
%
%

\IEEEPARstart{A}{lthough} the ¡°Memory Wall
\cite{Wulf95:MemoryWall}¡± problem has been raised for over one
decade, this trend remains in multicore era where both memory
latency and bandwidth become critical. Memory trace analysis is an
important technology for architecture research, system software
(i.e., OS, compiler) optimization, and application performance
improvements.

Regarding trace collection, Uhlig and Mudge
\cite{Uhlig97:MemorySimulation} proposed that an ideal memory trace
collector should be:
\begin{itemize}
  \item {\bf Complete:} Trace should include all memory references
made by OS, libraries and applications.

  \item {\bf Detail:} Trace should contain detail information to
distinguish one process¡¯ address space from others.

  \item {\bf Undistorted:} Trace should not include any additional
memory references and it should have no time dilation.

  \item {\bf Portable:} Trace can still be tracked when moving to
other machines with different configurations.

  \item {\bf Other characteristics:} An ideal trace collector should be
fast, inexpensive and easy to operate.
\end{itemize}

Memory trace can be collected in several ways which are either
hardware-based or software-based, such as software simulators,
binary instrumentation, hardware counters, hardware monitors, and
hardware emulators. Table \ref{Tab:memory_trace_sum} summarizes
these approaches. Although all approaches have their pros and cons,
hardware-snooping is relatively a more effective and efficient
approach to monitor and collect memory trace. Usually they are able
to collect undistorted and complete memory traces that include VMM,
OS, library and application. However, in contrast with
software-based approaches, there is a semantic gap between memory
traces collected by hardware-based approaches and high-level events,
such as kernel-user context switch , process/function/loop
execution, virtual address reference and I/O access.

\begin{table}
  \centering
  \caption{Summary of Memory Tracing Mechanism}\label{Tab:memory_trace_sum}
  \begin{tabular}{|l|c|c|c|c|c|}
    \hline
                & Simu- & Instr- & HW & HW~ & HW \\
                & late & ument & Cnt & Snoop & Emulate \\ \hline
    Complete    & \textasteriskcentered   & \textasteriskcentered  & x & $\surd$ & $\surd$ \\ \hline
    Detail      & $\surd$ & \textasteriskcentered & x & x & $\surd$ \\ \hline
    Undistorted & $\surd$ & x & $\surd$ & $\surd$ & x \\ \hline
    Portable    & $\surd$ & \textasteriskcentered & \textasteriskcentered & x & \textasteriskcentered \\ \hline
    Fast        & x & x & $\surd$ & $\surd$ & $\surd$ \\ \hline
    Inexpensive & $\surd$ & $\surd$ & $\surd$ & \textasteriskcentered & x \\ \hline
    \multicolumn{6}{c}{Note: ~~~~~~~~~~~$\surd$-Yes~~~~~~~~~~~  \textasteriskcentered-Maybe ~~~~~~~~~~~ x-No } \\
  \end{tabular}
\vspace{-0.2cm}\end{table}

To bridge the semantic gap, we propose a hybrid hardware/software
mechanism which is able to collect memory reference trace as well as
high-level event information. The mechanism integrates a flexible
software tracing-control mechanism into conventional
hardware-snooping mechanisms. A specific physical address region is
reserved as hardware components' configuration space which is
prohibited for any programs and OS modules except tracing-control
software components. When high-level events happen, the software
components inject a specific memory traces with semantic information
by referencing the pre-defined configuration space. Therefore,
hardware components can monitor and collect mixed traces which
contain both normal memory reference traces as well as high-level
event identifiers. In such a hybrid tracing mechanism, we are able
to analyze memory behaviors of specific events. For example, as
illustrated in section 6.3, we can distinguish I/O memory reference
from CPU memory reference. Moreover, the hybrid mechanism supports
various hardware-snooping methods, such as MemorIES
\cite{Nanda00:MemorIES} which snoops on the IBM¡¯s 6xx bus, PHA\$E
\cite{Chalainanont03:PHASE} and ACE \cite{Hong06:ACE} which snoop on
Intel's Front Side Bus (FSB) and our approach which snoops on memory
bus.

Based on this mechanism, we have designed and implemented a
prototype system called HMTT (Hybrid Memory Trace Tool) which adopts
a DIMM-snooping mechanism and a software-controlled trace mechanism
to inject semantic information into normal memory trace. Several new
techniques are proposed to overcome the system design challenges:
(1) To keep up with memory speeds, the DDR state machine
\cite{JEDEC} is simplified to match high speed, and large FIFOs are
added between the state machine and the trace transmitting logic to
handle occasional bursts. (2) To support flexible
software-controlled tracing, we develop a kernel module for an
uncachable memory region reservation. (3) To dump full mass traces,
we use a straightforward method to compress memory trace and adopt a
combination of Gigabit Ethernet and RAID to transfer and save the
compressed trace. Based on these primitive functions the HMTT system
provided, advanced functions can be designed, such as distinguishing
one process¡¯ address space from others, distinguishing I/O memory
references from CPU's. Comprehensive validations and evaluations
show that the HMTT system has both hardware's and software's
advantages. In summary, it has the following advantages:
\begin{itemize}
  \item {\bf Complete:} It is able to track complete memory reference trace
of real systems, including OS, VMMs, libraries, and applications.

  \item {\bf Detail:} The memory trace includes timestamp, r/w, and some semantic identifiers,
  e.g. process' pid, page table information, kernel entry/exit tags etc. It is
easy to differentiate processes' address spaces.

  \item {\bf Undistorted:} There are almost no additional references in most cases
  because the number of high-level events is much less than that of memory reference
  trace.

  \item {\bf Portability:} The hardware boards are plugged in DIMM slots
which are widely used. It is easy to port the monitoring system to
machines with different configurations (CPU, bus, memory etc.). The
software components can be run on various OS platforms, such as
Linux and Windows.

  \item {\bf Fast:} There is no slowdown when collecting memory trace for
analysis of L2/L3 cache, memory controller, DRAM performance and
power. The slowdown factor is about 10X$\sim$100X when cache is
disabled to collect whole trace.

  \item {\bf Inexpensive:} We have built the HMTT system, from
schematic, PCB design and FPGA logic to kernel modules, and analysis
programs. The implementation of hardware boards is simple and low
cost.

  \item {\bf Easy to operate:} It is easy to operate the HMTT system, because it provides several
toolkits for trace generation and analysis.
\end{itemize}

Using the HMTT system on X86/Linux platforms, we have investigated
the impact of OS on stream-based access and found that OS virtual
memory management can decrease stream accesses in view of memory
controller (or L2 Cache), by up to 30.2\% (301.apsi). We have found
that prefercher in memory controller cannot produce an expected
effect if not considering the multicore impact, because the
interference of memory accesses from multiple cores (i.e.,
processes/threads) is serious. We have also analyzed
characterization of DMA memory references and found that previously
proposed Direct Cache Access (DCA) scheme \cite{DCA:05Huggahalli}
have poor performance for disk-intensive applications because disk
I/O data is so large that it can cause serious cache interference.
In summary, the evaluations and case studies show the feasibility
and effectiveness of the hybrid hardware/software tracing mechanism
and techniques.

The rest of this paper is organized as follows. Section 2 introduces
semantic gap between memory trace and high-level events. Section 3
describes the proposed hybrid hardware/software tracing mechanism.
Section 4 presents design and implementation of the HMTT system and
section 5 discusses its verifications and evaluations. Section 6
presents several case studies of the HMTT system to show its
feasibility and effectiveness. Section 7 presents an overview of
related work. Finally, Section 8 summarizes the work.

\section{Semantic Gap Between Memory Trace and High-Level Events}

\begin{figure}[!t]
\centering
\includegraphics[width=3.2in]{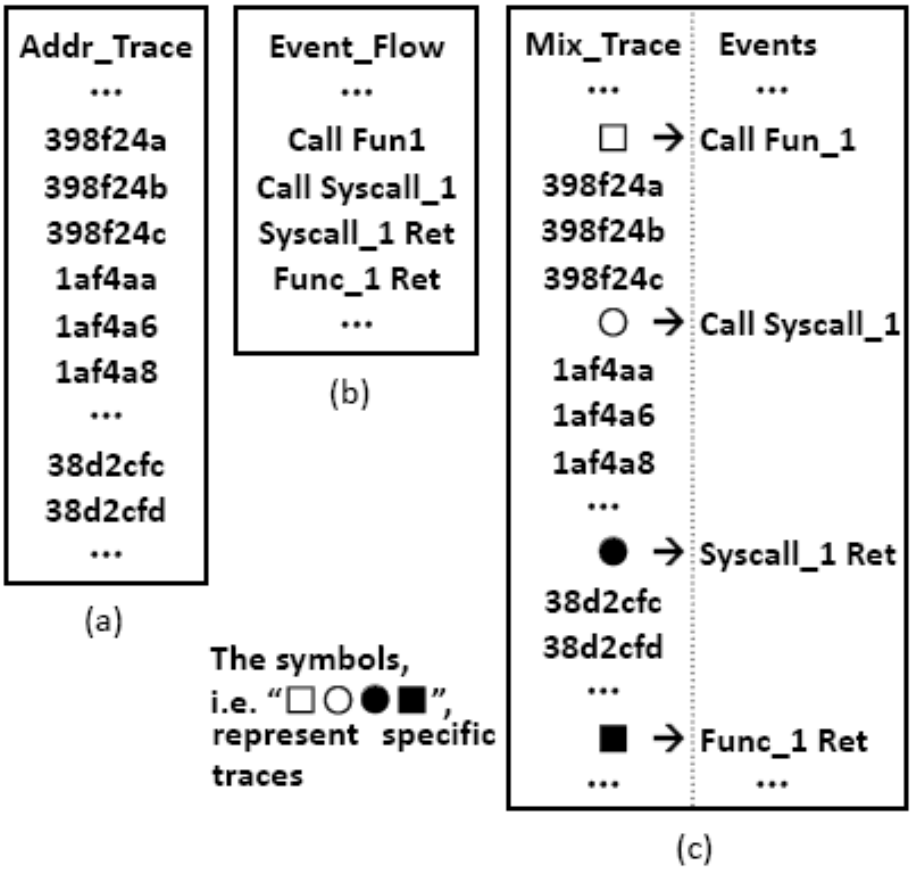}
\caption{The Semantic Gap between Memory Trace and High-Level
Events. (a) A conventional memory address trace; (b) A typical
high-level event flow; (c) The correlation of memory trace and
high-level events.} \label{fig:semantic_gap}
\vspace{-0.2cm}\end{figure}

Memory trace (or memory address trace) is a sequence of memory
references which are generated by executing load-store instructions.
Conceptually, memory trace mainly indicates instruction-level
architectural information. Figure \ref{fig:semantic_gap}(a) shows a
conventional memory trace (in which timestamp, read/write and other
information have already been removed). Since trace-driven
simulation is an important approach to evaluate memory systems and
has been used for decades \cite{Uhlig97:MemorySimulation}, this kind
of memory trace has played a significant role in advancing memory
system performance. As described in the introduction, memory trace
can be collected in various ways among which hardware-snooping is
relatively a more effective and efficient approach. Usually they are
able to collect undistorted and complete memory traces that include
VMM, OS, library and application. Nevertheless, those memory traces
mainly reflect low-level (machine-level) information which is
obscure for most people.

From the perspective of system level, a computer system generates
various events, such as context switch, function call, syscall and
I/O request. Figure \ref{fig:semantic_gap}(b) illustrates a typical
event flow. To capture high-level event flow, one may instrument
source code or binary at points of these events manually or
automatically. In contrast with memory trace, those events are at
higher levels and contain more semantic information which people can
understand more easily. However, only given high-level events, it is
usually insufficient to analyze system's performance and behaviors
in depth.

Based on the above observations, we can conclude that there is a
semantic gap between conventional memory traces and high-level
events. If they can correlate to each other, as shown in Figure
\ref{fig:semantic_gap}(c), it should be significantly helpful for
both of low-level (memory trace) and high-level (system or program
events) analysis. For example, for architecture and system, one can
distinguish I/O memory references from CPU memory references or
analyze memory access pattern of a certain syscall, function and
loop and so on. For software engineering, memory access information
can be gathered for performance analysis to determine which sections
of a program to optimize to increase its speed or decrease its
memory requirement.

However, prevalent trace tools can only collect either memory trace
or function call graph and OS event. Some hardware monitors are only
capable of collecting whole memory requests by snooping on memory
bus, such as MemorIES \cite{Nanda00:MemorIES}, PHA\$E
\cite{Chalainanont03:PHASE} and ACE \cite{Hong06:ACE}. For high
level events, gprof can only provide call graph, and Linux Trace
Toolkit \cite{LTT} and Lockmeter \cite{Lockmeter:00Bryant} focus on
collecting operating system events, however, with a substantial
amount of overhead. In addition, by instrumenting the target program
with additional instructions, some instrument tools such as ATOM
\cite{Srivastava94:ATOM}, Pin \cite{PIN-tool}, Valgrind \cite{Valgrind}
are capable of collect more information, e.g., memory trace,
function call graph. However, complicated instrumenting the program
can cause changes of the performance of program, inducing inaccurate
results and bugs. Instrumenting can also slow down the target
program as more specific information is collected. Moreover, it is
hard to instrument virtual machine monitor and operating system.

In summary, there is a semantic gap between conventional memory
traces and high-level events but almost none of the existing tools
are capable of bridging the gap effectively.

\section{A Hybrid Hardware/Software Tracing Mechanism}
To address the semantic gap, we propose a hybrid hardware/software
mechanism which is able to collect memory reference trace as well as
high-level event information simultaneously.

As shown in Figure \ref{fig:semantic_gap}(c), in order to
efficiently collect such correlated memory reference and event
trace, the hybrid tracing mechanism consists of three key parts
which we will discuss in the following subsections.

\subsection{Hardware Snooping}
Hardware snooping is an efficient approach to collect memory
reference via snooping on system bus or memory bus. It is able to
collect complete memory traces including VMM, OS, library and
application without time and space distortions. It should be noted
that hardware snooping approach mainly collects off-chip traffics.
Nevertheless, there are at least two ways to alleviate this negative
influence while one needs all memory references generated by
load/store unites within a chip: (1) Mapping program's virtual
address regions to physical memory with uncachable attribution. This
can be done by a slight modification for OS memory management or a
command to reconfigure processor's Memory Type Range Registers
(MTRR). (2) Enabling/disabling cache dynamically. To achieve such a
goal, we can set cache control registers many processors provided
(e.g., X86's CR0) when entering a certain code section. These cache
control approaches may cause slowdown of 10X, still being
competitive while comparing to other software tracing approaches.

\subsection{Configuration Space}
Usually it is difficult for software to control and synchronize with
hardware snooping devices, because the devices are usually
independent of target traced machine.

\begin{figure}[!t]
\centering
\includegraphics[width=2.5in]{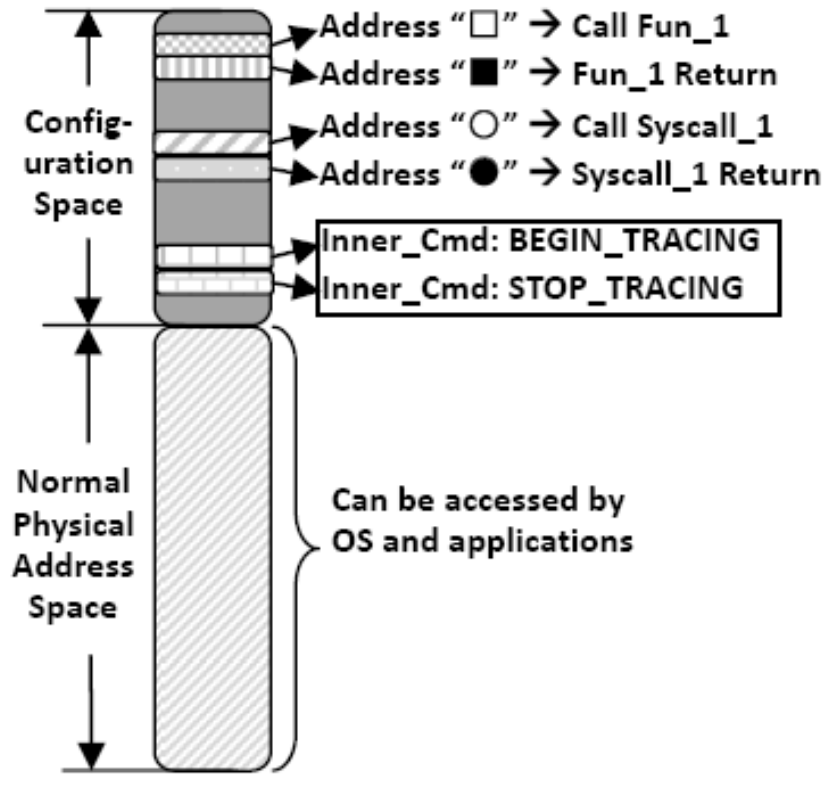}
\caption{System's physical address space is divided into two parts:
(1) a normal space which can be accessed by OS and applications and
(2) a specific space which is reserved as hardware snooping device's
configuration space. The addresses within the configuration space
represent either inner commands or high-level events.}
\label{fig:config_space} \vspace{-0.2cm}\end{figure}

We address this problem by introducing a specific physical address
region reserved as hardware device's configuration space which is
prohibited for any programs and OS modules except tracing-control
software, as illustrated in Figure \ref{fig:config_space}. The
addresses within the configuration space can be predefined as
hardware device's inner commands, such as \emph{BEGIN\_TRACING,
STOP\_TRACING, INSERT\_ONE\_SPECIFIC\_TRACE}. They can also
represent high-level events, such as function call and syscall
return.

\subsection{Low Overhead Tracing-Control Software}
Based on the configuration space, a low overhead tracing-control
software mechanism can be integrated into a conventional
hardware-snooping mechanism.

The tracing-control software has two functions. First, it is able to
control hardware snooping device. When the tracing-control software
generates a memory reference to a specific address in the
configuration space, the hardware device captures the specific
address which is predefined as an inner command, such as
\emph{BEGIN\_TRACING} or \emph{STOP\_TRACING}. Then the hardware
device performs corresponding operations according to the inner
command. Second, the software can make hardware snooping device
synchronize with high-level events. When those events occur, the
tracing-control software generates specific memory references to the
configuration space in which different addresses represent different
high-level events. In this way, the hardware device is able to
collect mixed traces as shown in the left side of Figure
\ref{fig:semantic_gap}(c), including both normal reference and
specific reference.

\begin{figure}[!t]
\centering
\includegraphics[width=3.2in]{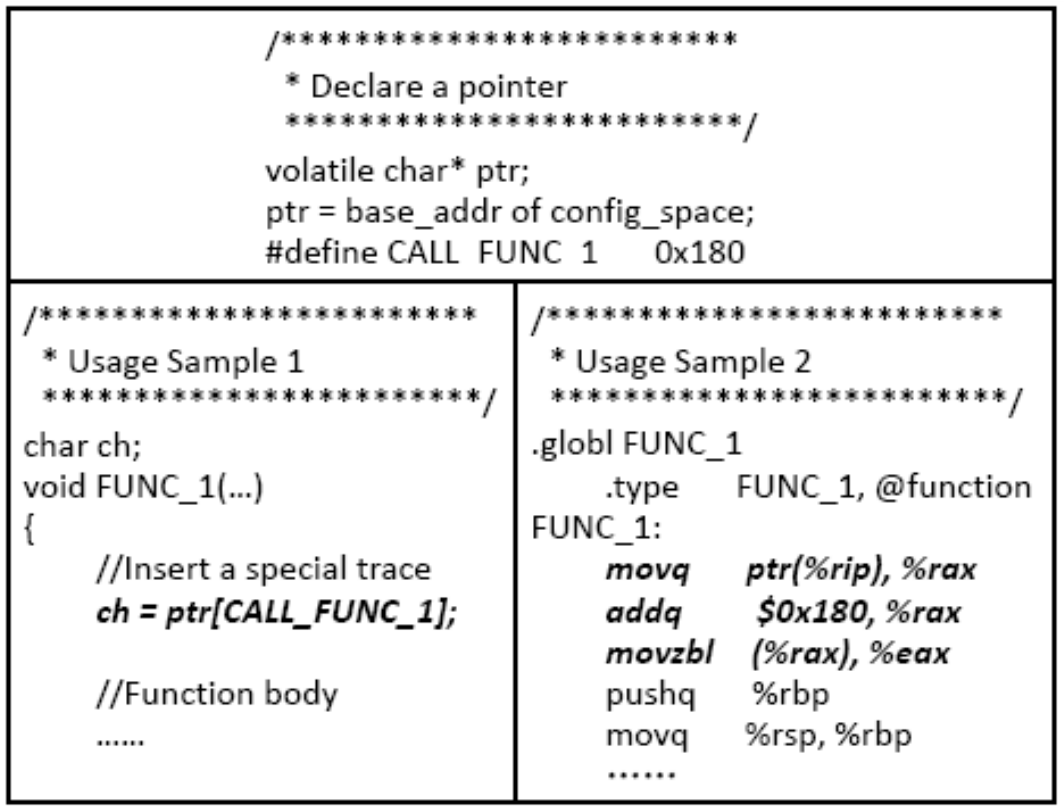}
\caption{Two Samples of Tracing-Control Software. Programer and
compiler can instrument programs with the codes manually or
automatically. These codes contain high-level events information and
will issue specific memory references to the configuration space of
hardware snoop device.} \label{fig:control_software}
\vspace{-0.2cm}\end{figure}

Since hardware snooping can be controlled by only one memory
reference, this tracing-control software mechanism is extremely
low-overhead. The design and implementation of the software are
quite simple. Figure \ref{fig:control_software} illustrates a sample
of tracing-control software. It includes two phases. In phase one, a
pointer {\bf \emph{ptr}} is defined and assigned with base address
of the configuration space. In phase two, programs can be
instrumented with the statement {\bf \emph{"ch =
ptr[EVENT\_OFFSET];"}} to insert specific references into normal
trace. Further, in order to reduce substantial negative influence of
source code instrumentation, instructions can be directly inserted
into an assembler program, as the second sample shown in Figure
\ref{fig:control_software}.

With this hybrid tracing mechanism, we are able to analyze memory
behaviors of a certain event. For example, as illustrated in section
6, we can instrument device drivers to distinguish I/O memory
reference from CPU memory reference. Further, the tracing mechanism
can be configured to only collect high-level events with very low
overhead, i.e., only collect the right side of Figure
\ref{fig:semantic_gap}(c). In addition, the hybrid mechanism
supports various hardware-snooping methods, such as MemorIES
\cite{Nanda00:MemorIES} which snoops on the IBM¡¯s 6xx bus, PHA\$E
\cite{Chalainanont03:PHASE} and ACE \cite{Hong06:ACE} which snoop on
Intel's Front Side Bus (FSB) and our prototype system HMTT which
snoops on memory bus.

\section{Design and Implementation of HMTT Tracing System}
Based on the hybrid hardware/software tracing mechanism, we have
designed and implemented a prototype system called HMTT (Hybrid
Memory Trace Tool). The HMTT system adopts a DIMM-snooping mechanism
that uses hardware boards plugged in DIMM slots to snoop on memory
bus. We will introduce design and implementation of the HMTT system
in detail in the following subsections.

\subsection{Top-Level Design}
\begin{figure}[!t]
\centering
\includegraphics[width=2.5in]{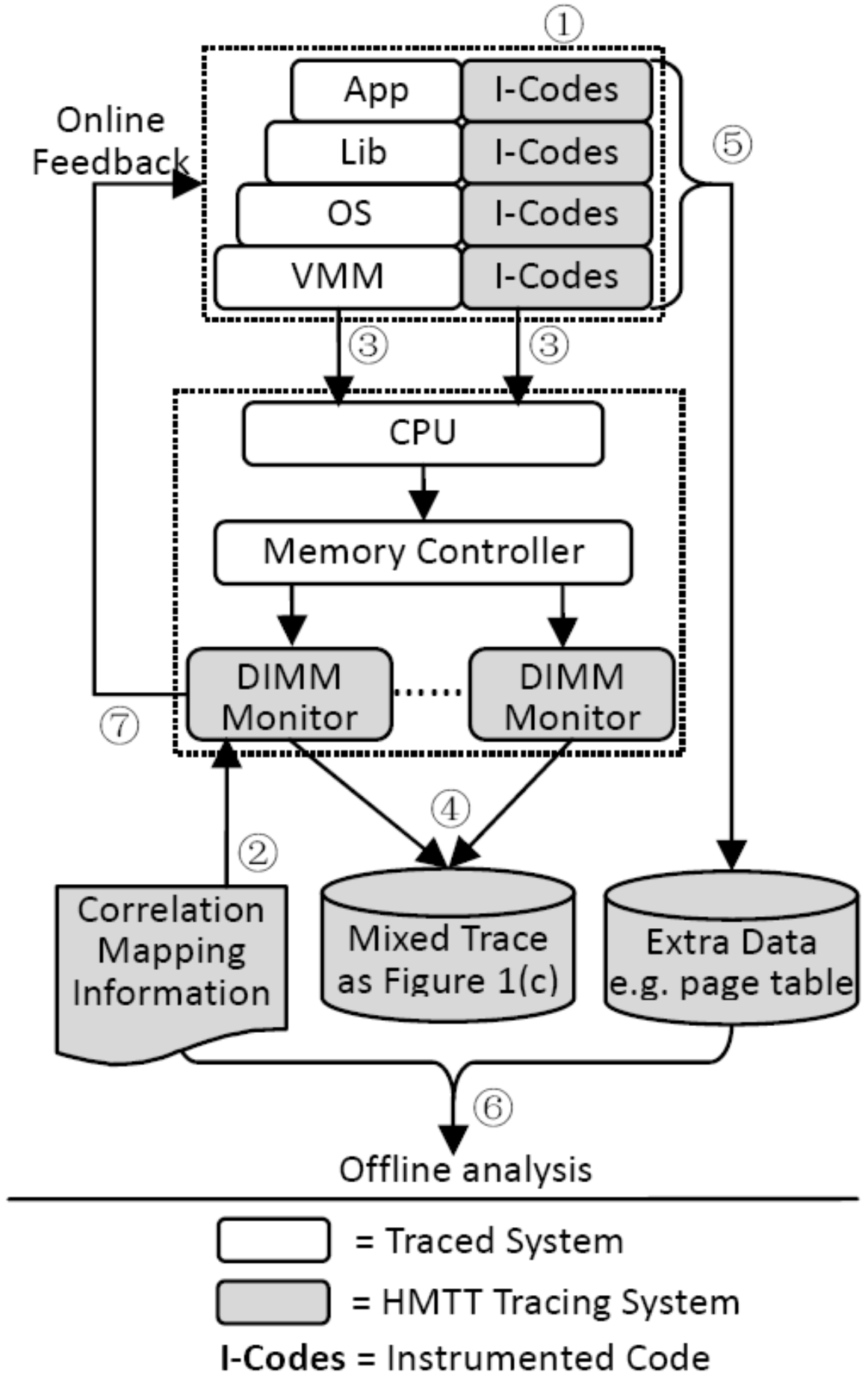}
\caption{The Framework of HMTT Tracing System. It contains seven
procedures: \textcircled{1} instrument target program manually or
automatically to generate I-Codes and correlation mapping
information (\textcircled{2}). \textcircled{3} generate memory
references; \textcircled{4} hardware snooping devices collect and
dump mixed trace to storage. \textcircled{5} I-Codes generate extra
data if necessary. \textcircled{6} replay trace for offline
analysis. \textcircled{7} hardware snooping devices can perform
online analysis for feedback optimization.} \label{fig:hmtt_design}
\vspace{-0.2cm}\end{figure}

At the top-level, the HMTT tracing system mainly consists of seven
procedures for memory trace tracking and replaying. Figure
\ref{fig:hmtt_design} shows the system framework and the seven
procedures.

From Figure \ref{fig:hmtt_design}, the first step for mixed trace
collection is instrumenting target programs (i.e., application,
library, OS and VMM) with I-Codes by hand and by scripts or
compilers (\textcircled{1}). The I-Codes inserted at the points
where high-level events occur will generate specific memory
references (\textcircled{3}) and some extra data such as page table,
I/O request (\textcircled{5}). Note that the mapping information of
correlated memory trace and high-level events (\textcircled{2}) is
also an output of the instrumenting operations.

For hardware parts, the HMTT system uses several hardware
DIMM-monitor boards plugged into DIMM slots of a traced machine. The
main memories of the traced system are plugged into the DIMM slots
integrated on the hardware monitoring boards (see Figure
\ref{fig:hmtt_v2}). The boards monitor all memory commands via DIMM
slots (\textcircled{4}). An on-board FPGA converts the commands into
memory traces in this format $<$address, r/w, timestamp$>$. Each
hardware monitor board generates trace separately and sends the
trace to its corresponding receiver via Gigabit Ethernet or
PCI-Express interface (see \textcircled{4}). With synchronized
timestamps, the separated traces can be merged into a total mixed
trace.

If necessary, the I-Codes can track and collect some additional data
to aid memory trace analysis. For example, page table information
can be used to reconstruct physical-to-virtual mapping relationship
to help analyze process' virtual address (\textcircled{6}). I/O
request information collected from device drivers can be used to
distinguish I/O memory references from CPU memory references.
Further, the on-board FPGA can perform online analysis and send
feedbacks to OS for online optimization via interrupt.

We need to address several challenges to design this system, such as
how to make hardware snooping devices keep up with memory speeds,
how to design configuration space for hardware devices, how to
control tracing by software and how to dump and replay massive
trace. We will elaborate on our solutions in the following
subsections.

\subsection{Hardware Snooping Device}

\subsubsection{Keeping up with Memory Speeds}

\begin{figure}[!t]
\centering
\includegraphics[width=2.5in]{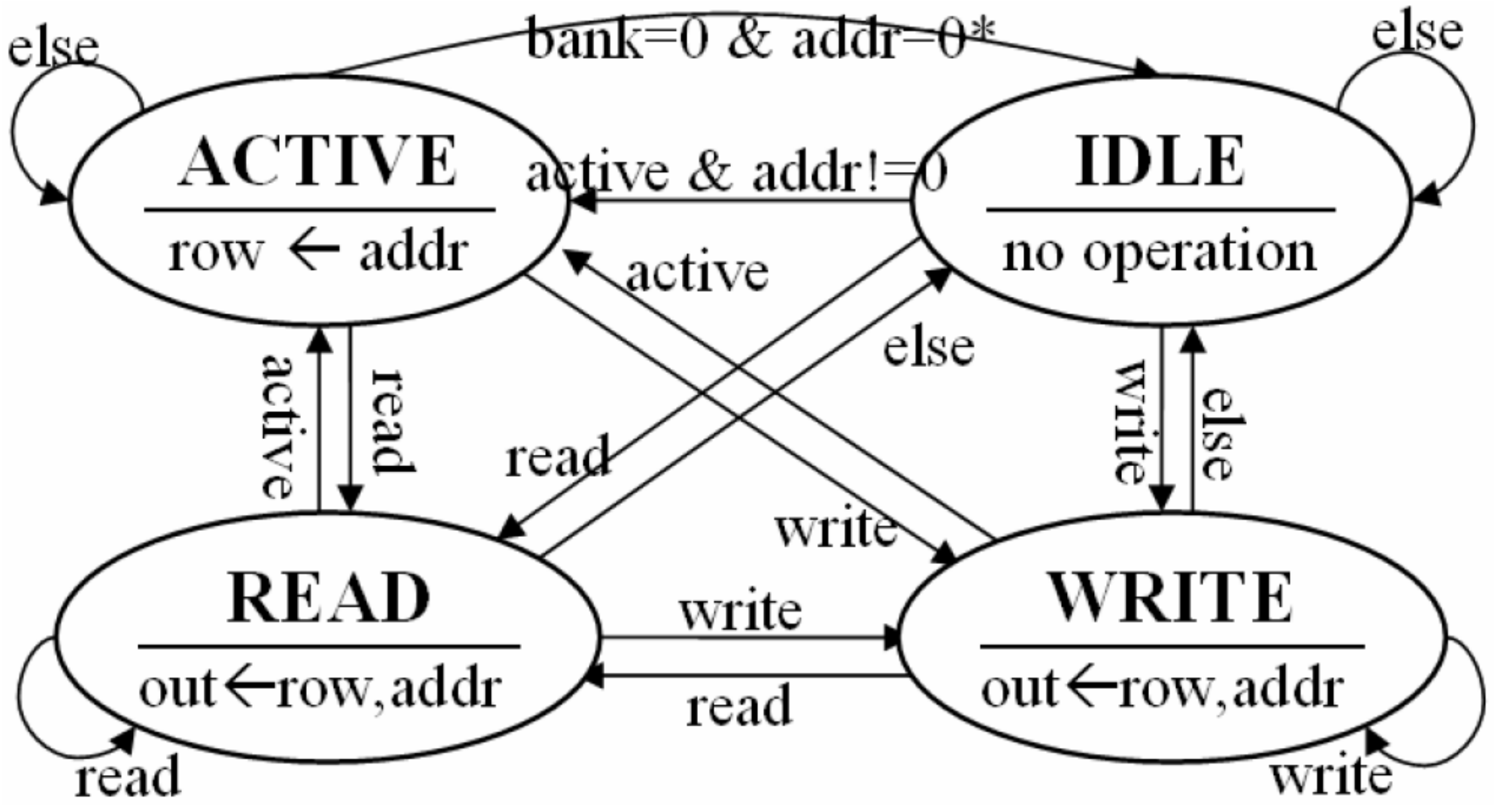}
\caption{Simplified State Machine. To keep up with memory speeds,
the standard DDR state machine \cite{JEDEC} is simplified to match
high speed. (* - Note that \emph{"addr"} is used to filiter
specifical addresses of configuration space.)}
\label{fig:hmtt_state_machine} \vspace{-0.2cm}\end{figure}

Fast and efficient control logic is required to keep up with memory
speeds because of high memory frequency and multi-bank technologies.
Since only memory address is indispensable for tracking trace, we
can only monitor DDR commands at half memory data frequency. For
example, if use a DDR2-533MHz memory, the control logic can operate
at a frequency of only 266MHz, at which most advanced FPGAs can
work.

To interpret the two-phase read/write operations, the DDR SDRAM
specification \cite{JEDEC} defines seven commands and a state
machine which has more than twelve states. Commercial memory
controllers integrate even more complex state machines which cost
both time and money to implement and validate. Nevertheless, we find
that only three commands, i.e. ACTIVE, READ and WRITE, are necessary
for extracting memory reference address. Thus, we design a
simplified state machine to interpret the two-phase operations for
one memory bank. Figure \ref{fig:hmtt_state_machine} shows the
simplified state machine. It has only four states and performs state
transitions based on the three commands. The state machine is so
simplified that the implementation in a common FPGA is able to work
at a high frequency. Our experiments show that the state machine
implemented in a Xilinx Virtex II Pro FPGA is able to work at a
frequency of over 300MHz.

On the other hand, applications will generate occasional bursts
which may induce dropping trace. A large FIFO between the state
machine and trace transmitting logic is provided to solve this
problem. In the HMTT system, we have verified that a 16K entries
FIFO is sufficient to match the state machine for a combination of
DDR-200MHz and a transmission bandwidth of 1Gbps as well as a
combination of DDR2-400MHz and a bandwidth of 3Gpbs. For a higher
memory frequency (e.g., DDR2-800MHz), we can adopt some alternative
transmission technologies, such as PCI-E which can provide
bandwidths of over 8Gbps.

\subsubsection{Design of Hardware Device}
The HMTT system consists of a Memory Trace Board (MTB), which is a
hardware monitor board wrapping a normal memory and itself plugged
in a DIMM slot (see Figure \ref{fig:hmtt_v2}). The MTB monitors
memory command signals which are sent to DDR SDRAM from memory
controller. It captures the DDR commands, and forwards them to the
simplified DDR state machine (described in the last subsection). The
output of state machine is a tuple $<$address, r/w, duration$>$.
These raw traces can be sent out directly via GE/PCIE or buffered
for online analysis.

\begin{figure}[!t]
\centering
\includegraphics[width=3in]{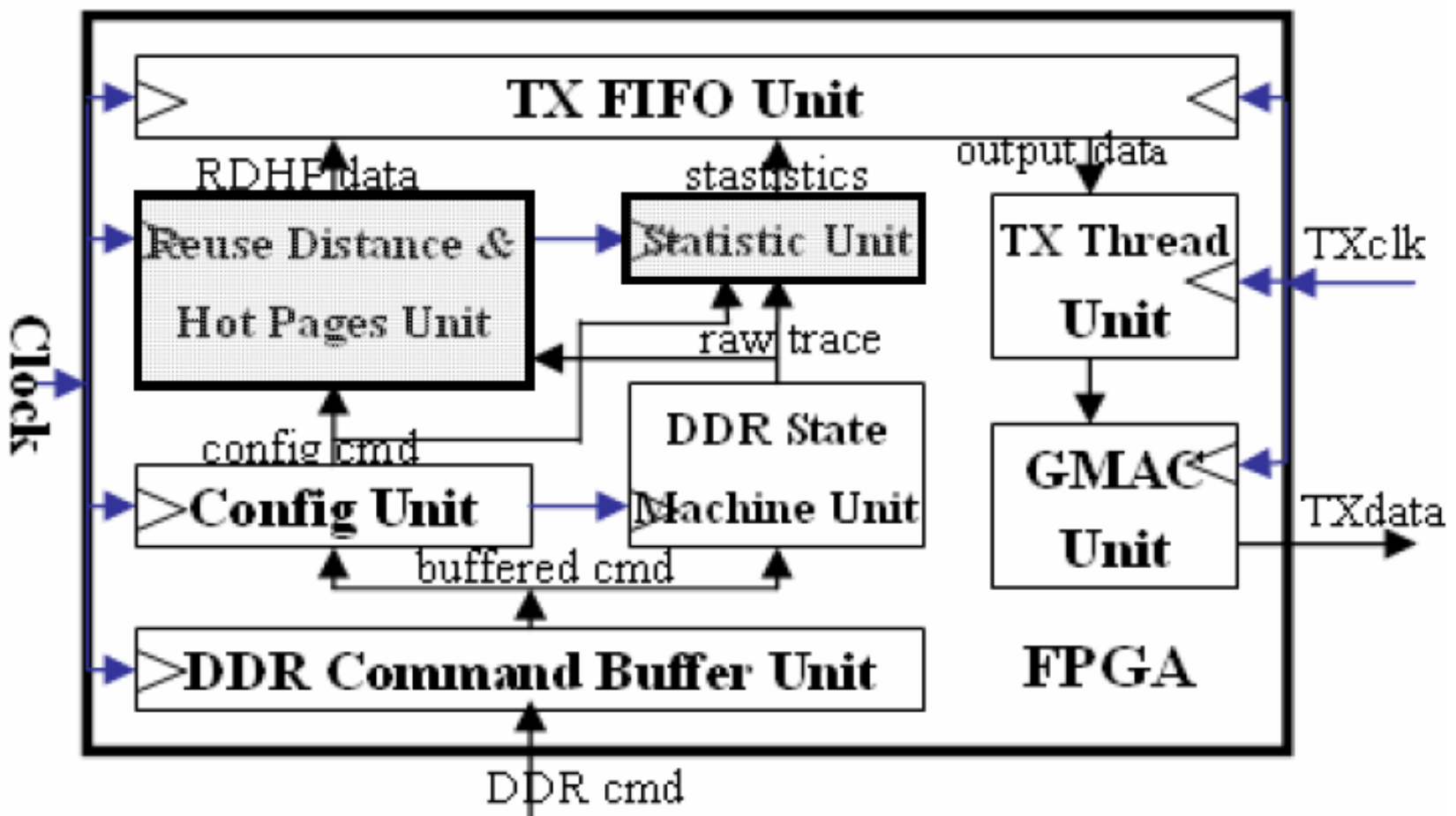}
\caption{The FPGA Physical Block Diagram. It contains eight logic
units. Note that the \emph{"Statistic Unit"} and \emph{"Reuse
Distance \& Hot Pages Unit"} are optional for online analysis.}
\label{fig:HMTT_FPGA} \vspace{-0.2cm}\end{figure}

There is an FPGA on the MTB. Figure \ref{fig:HMTT_FPGA} shows the
physical block diagram of the FPGA. It contains eight logic units.
The DDR Command Buffer Unit (DCBU) captures and buffers DDR
commands. Then the buffered commands are forwarded to the Config
Unit and the DDR State Machine Unit. The Config Unit (CU) translates
a specific address into inner-commands, and then controls MTB to
perform corresponding operations, such as switching work mode,
inserting synchronization tags to trace. The DDR State Machine Unit
(DSMU) interprets two-phase interleaved multi-bank DDR commands to a
format of $<$address, r/w, duration$>$. Then the trace will be
delivered to the TX FIFO Unit (TFU) and be sent out via GE. The FPGA
is reconfigurable to support two optional units: the Statistic Unit
(SU) and Reuse Distance \& Hot Pages Unit (RDHPU).

The Statistic Unit is able to do statistic of various memory events
in different intervals (1us $\sim$ 1s), such as memory bandwidth,
bank behavior, and address bits change. The RDHPU is able to
calculate page¡¯s reuse distance and collect hot pages. The RDHPU¡¯s
kernel is a 128-length LRU stack which is implemented in an enhanced
systolic array proposed by J.P. Grossman \cite{Grossman02:Systolic}.
The output of these statistic unit can be sent out or used for
online feedback optimization. To keep up with memory speeds, the DDR
State Machine Unit adopts the simplified state machine described in
the last subsection. The TX FIFO Unit contains a 16K entries FIFO
between the state machine and the trace transmitting logic.

\subsection{Design of Configuration Space}
As described before, we adopt a configuration space mechanism to
address the challenge of making software control hardware snooping
devices. Figure \ref{fig:config_space} has illustrated the principle
scheme of this mechanism where a specific physical address region is
reserved for configuration space. Further, the right part of Figure
\ref{fig:hmtt_config} illustrates the details of HMTT's
configuration space. The addresses of the space are defined as
either HMTT's inner commands (e.g., \emph{BEGIN\_TRACING,
STOP\_TRACING, INSERT\_ONE\_SPECIFIC\_TRACE}) or user-defined
high-level events (from address 0x1000). Note that the difference of
two contiguous defined addresses relies on block size of processor's
last level cache whose size is 64 (0x40) bytes in our cases.

\subsection{Tracing-Control Software}
\begin{figure}[!t]
\centering
\includegraphics[width=3.2in]{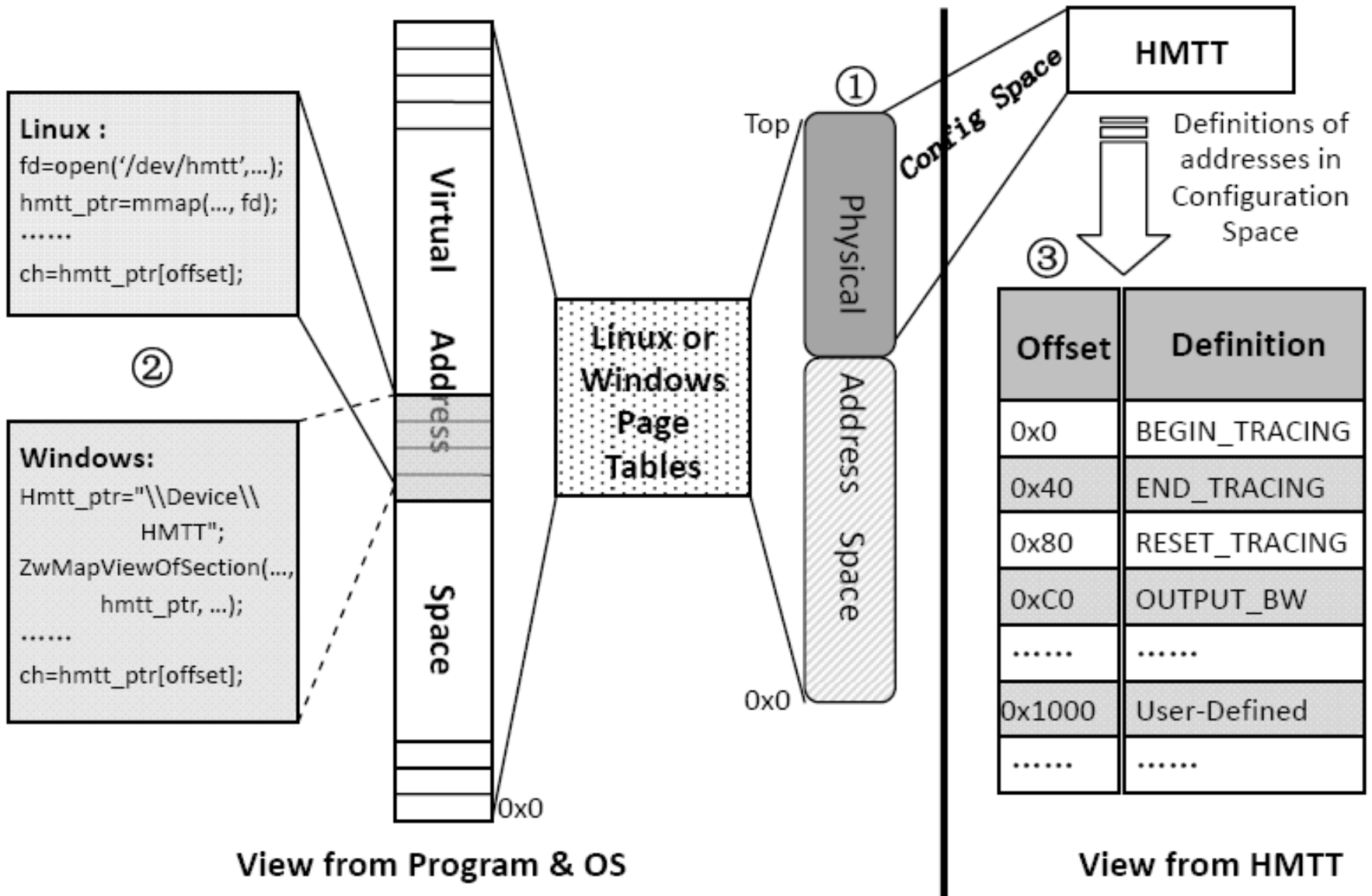}
\caption{The Software Tracing-Control Mechanism. \textcircled{1} The
physical memory is reserved as HMTT's configuration space.
\textcircled{2} User programs map the space into their virtual
address space. \textcircled{3} There are predefined inner commands
in the configuration space.} \label{fig:hmtt_config}
\vspace{-0.2cm}\end{figure}

Figure \ref{fig:control_software} has already illustrated two
samples of tracing-control software. In this subsection, we will
present details of software implementations. As shown in Figure
\ref{fig:hmtt_config}, the tracing-control software can run on both
Linux and Windows platforms. At phase \textcircled{1}, the top
several megabytes (e.g., 8MB) physical memory is reserved as the
HMTT's configuration space when Linux or Windows boot. This can be
done by modifying the parameter in grub (i.e., \emph{mem}) or
boot.ini (i.e., \emph{maxmem}). Thus, access to the configuration
space is prohibited for any programs and OS modules. At phase
\textcircled{2}, a kernel module is introduced to map the reserved
configuration space as a user-defined device, called /dev/hmtt for
Linux or $\backslash$$\backslash$Device$\backslash$$\backslash$HMTT
for Windows. Then user programs can map /dev/hmtt (Linux) or
$\backslash$$\backslash$Device$\backslash$$\backslash$HMTT (Windows)
into their virtual address spaces so that they can access the HMTT's
configuration space directly. At phase \textcircled{3}, the HMTT's
\emph{Config Unit} will identify the predefined addresses and
translate them into inner-commands to control the HMTT system. For
example, the inner-command \emph{END\_TRACING} is defined as one
memory read reference on the offset of 0x40 in the configuration
space.

\subsection{Trace Dumping and Replay}
Usually, memory reference traces are generated at very high speed.
Our experiments show that most applications generate memory trace at
bandwidths of more than 30MB/s even when utilize the DDR-200MHz
memory. Moreover, the high frequency of the DDR2/DDR3 memory and the
prevalent multi-channel memory technology further increase trace
data generation bandwidth, up to 100X MB/s. Our efforts consist of
two aspects.

First, we apply several straightforward compress methods to reduce
the memory trace generation and transmission bandwidth. While memory
works in burst mode \cite{JEDEC}, we only need to track the first
memory address of a contiguous addresses pattern. For example, when
the burst length is equal to four, the latter three addresses of a
4-length addresses pattern can be ignored. Trace format is usually
defined as $<$address, r/w, timestamp$>$ which needs at least
6$\sim$8 bytes to store and transmit. We find that the high bits of
the difference of timestamps in two adjacent traces are always 0s at
most time. We use \emph{duration (= $timestamp_{n}$ -
$timestamp_{n-1}$)} to replace timestamp in the trace format. This
differencing method reduces the duration bits to ensure one trace to
be stored and transmitted in 4 bytes. However, the duration may
overflow. We define a specific format $<$special\_identifier,
duration\_high\_bits$>$ to handle the overflows. Then, the
timestamps can be calculated in the trace replay phase. The
straightforward compress methods substantially reduce trace
generation and transmission bandwidth.

Second, the experimental results show that trace generation
bandwidth is still high with the above compressions. In the
procedure \textcircled{4} shown in figure \ref{fig:hmtt_design}, we
adopt multiple Gigabit Ethernets (GE) and RAIDs to send and receive
memory traces respectively (in fact, multiple GEs can be replaced by
one PCIE interface). In this way, all traces are received and stored
in RAID storages (the details about trace generation and
transmission bandwidth will be discussed in the next section). Each
GE sends trace respectively, so the separated traces need to be
merged when replay. We assign each trace its own timestamp which is
synchronized within FPGA. Then the trace merge operation is
simplified to be a merge sort problem.

In summary, the HMTT system adopts a combination of the
straightforward compressions, the GE-RAID approach and the trace
merge procedure to dump massive traces. In our experiments, we use a
machine with several Intel E1000 GE NICs to receive memory trace.
These techniques are scalable for higher trace generation bandwidth.

\subsection{Other Design Issues}
There are several other design issues of the HMTT system, such as
collecting extra kernel information, controlling cache dynamically.

{\bf Assistant Kernel Module:} We introduce an assistant kernel
module to help collect kernel information, such as page table, I/O
requests. On Linux platform, the assistant kernel module provides an
\emph{hmtt\_printk} routine which can be called at any place from
the kernel. Unlike Linux kernel¡¯s printk, the hmtt\_printk routine
supports large buffers and user-defined data format, like some
popular kernel log tools, such as LTTng \cite{LTT}. The assistant
kernel module requires a kernel buffer to store kernel collected
information. Usually, this buffer is quite small. For example, our
experiments show that the size of a buffer for all page tables is
only about 0.5\% of total system memory.

{\bf Dynamical Cache-Enable Control:} Hardware snooping approach
collects off-chip traffics. We adopt a dynamically
enabling/disabling cache approach to collect all memory references.
On X86 platforms, we introduce a kernel module to set the
Cache-Disable bit (bit30) of CR0 register to control cache when
entering or exiting a certain code section. This cache control
approach may cause slowdown of 10X, still being competitive while
comparing to other software tracing approaches. In addition, the X86
processor's Memory Type Range Registers (MTRR) can also be used for
managing cachable attribution.

\subsection{Put It All Together}
So far, we have described a number of design issues of the HMTT
system, including hardware snooping device, configuration space,
tracing-control software, trace dumping and replay, assistant kernel
module, dynamically cache enabling and so on.

\begin{figure}[!t]
\centering
\includegraphics[width=3in]{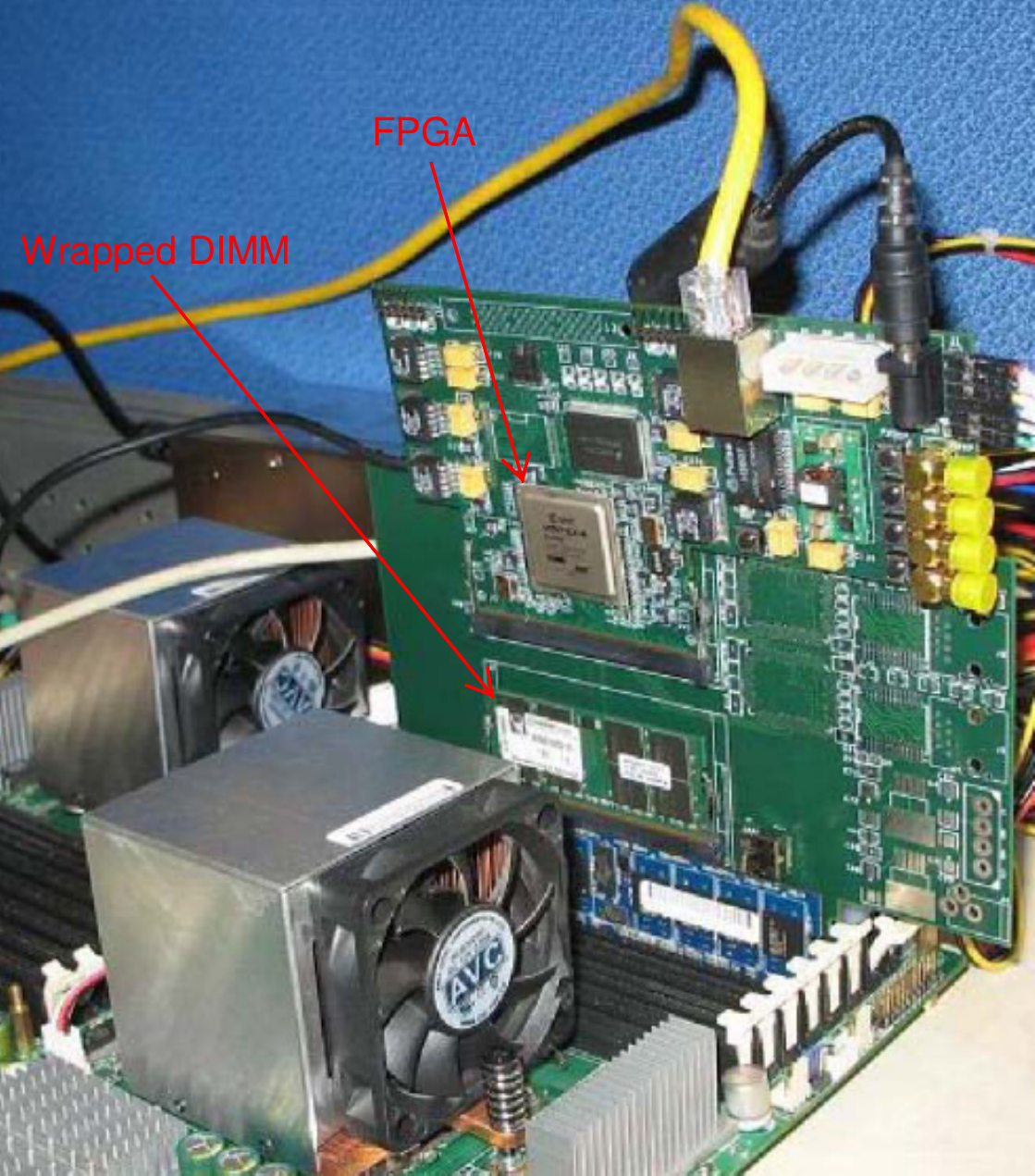}
\caption{The HMTT Tracing System. It is plugged into a DIMM slot of
the traced machine. The main memory of the traced system is plugged
into the DIMM slot integrated on the HMTT system.}
\label{fig:hmtt_v2} \vspace{-0.2cm}\end{figure}

Figure \ref{fig:hmtt_v2} illustrates the real HMTT system which is
working on an AMD server machine. Currently, the HMTT system
supports DDR-200MHz and DDR2-400MHz and will support
DDR2/DDR3-800MHz in the near future. We have tested the
tracing-control software on various Linux kernels (2.6.14 $\sim$
2.6.27). The software can be ported to Windows platform easily. We
have also developed an assistant kernel module to collect page table
and DMA requests currently (we will describe them in detail later).
Besides, we have developed a toolkit for trace replay and analysis.

\section{Evaluations of the HMTT System}
\subsection{Verification and Evaluation}
We have done a lot of verification and evaluation work. The HMTT
system is verified in four aspects, including physical address,
comparison with performance counter, software components and
synchronization of hardware device and software. We have also
evaluated the overheads of the HMTT system, such as trace bandwidth,
trace size, additional memory references, execution time of I-Codes
and kernel buffer for collecting extra kernel data. The work show
that the HMTT system is an effective and efficient memory tracing
system. (More details are shown in APPENDIX A.)

\subsection{Discussion}

\subsubsection{Limitation}
 It is important to note that the monitoring mechanism
can not distinguish the prefetch commands.

Regarding the impact of prefetch on memory trace, it has both up
side and down side. The up side is that we can get real memory
accesses trace to main memory, which can benefit research on main
memory side (such as memory thermal model research
\cite{Lin07:ThermalDRAM}). The down side is that it is hard to
differentiate the prefetch memory access and on-demand memory
accesses. Regarding prefetch, caches could generate speculative
operations. However, they do not influence memory behaviors
significantly. Most memory controllers do not have complex prefetch
unit, although several related efforts have been made, such as
Impluse project \cite{Zhang01:Impulse}, proposed region prefetcher
\cite{Wang03:RegionPrefetch}, and the stream prefetcher in memory
controller \cite{Hur06:MemoryPrefetch}. Thus, it is not a critical
weakness of our monitoring system. It is to be noted that all
hardware monitors also have the same limitation, prefetching from
various levels of the memory hierarchy.

\subsubsection{Combination With Other Tools}
As a new tool, HMTT system is a complementary tool to binary
instrumentation and full system simulation with software rather than
a thorough substitution. Since it is running in real-time and in
real systems, the combination with different techniques would be
more efficient for architecture and application research.

{\bf Combination with simulators:} to combine with simulators, the
HMTT system can be used to collect trace from real systems,
including multicore platform. Then, the trace is analyzed for
finding new insights. Some new optimization mechanisms based on new
insights can be evaluated by simulators.

{\bf Combination with binary instrumentation:} In fact, the
tracing-control software is an instance of the combination of
hardware snooping and source code instrumentation. Further, we can
adopt binary instrumentation to insert tracing-control codes into
binary files to identify functions/loops/blocks. In addition, with
compiler-provided symbol table, the virtual-address trace can be
used for semantic analysis.

\section{Case Studies}
In this section, we will present several case studies on two
different platforms, an Intel Celeron machine and an AMD Opteron
machine respectively. The case studies are: (1) OS impact on
stream-based access; (2) multicore impact on memory controller; (3)
characterization of DMA memory reference.

We have performed experiments on two different machines listed in
Table \ref{Tab:Platform_Param}. It should be noted that the HMTT
system can be ported to various platforms, including multicore
platforms, because it mainly depends on DIMM. We have studied memory
behaviors of three classes of benchmarks including (See Table
\ref{Tab:Platform_Param}): computing intensive applications (SEPC
CPU2006 and SPEC CPU2000 with reference input sets), OS intensive
applications (OpenOffice and Realplayer), Java Virtual Machine
applications (SPECjbb 2005), and I/O-intensive applications
(File-Copy, SPECWeb 2005 and TPC-H).

\begin{table}[htbp]
\centering
  \caption{Experimental Machines and Applications}\label{Tab:Platform_Param}
\begin{tabular}{|c||l|l|c|c|}
  \hline
        & {\bf Machine 1}     & {\bf Machine 2} \\
  \hline
  \hline
  CPU        & Intel Celeron 2.0GHz &  AMD Opteron   \\
             &                      &  Dual Core 1.8GHz \\\hline
  L1 I-Cache & 12K,6$\mu$op/Line   &  64KB,64B/Line  \\ \hline
  L1 D-Cache & 8KB,4-Way,64B/Line  &  64KB,64B/Line  \\ \hline
  L2 Cache   & 128KB,2-Way,64B/Line &  1MB,16-Way,64B/Line  \\ \hline
  Memory     & Intel 845PE  & Integerated   \\
  Controller &              &   \\ \hline
  DRAM       & 512MB,DDR-200 &  4GB,DDR-200  \\
             &               &  Dual-Channel  \\ \hline
  Hardware   & None          &  Yes  \\
  Counter    &               &       \\ \hline
  Hardware   & None          & Sequential Prefetcher  \\
  Prefetcher &               & in Memory Controller  \\ \hline
  OS         & Fedora Core 4(2.6.14) & Fedora 7(2.6.18)  \\
  \hline
             & 1.SPEC CPU 2000 &   1. SPEC CPU 2006  \\
  Application     & 2.SPECjbb 2005 &    2. File-Copy: 400MB  \\
             & 3.OpenOffice: 25MB slide & 3. SPECWeb 2005  \\
             & 4.RealPlayer: 10m video & 4. Oracle + TPC-H  \\
  \hline
\end{tabular}
\vspace{-0.2cm}\end{table}

\subsection{OS Impact on Stream-Based Access}
Stream-based memory accesses, also called fixed-stride access, can
be used by many optimization approaches, such as prefetching and
vector loads. Here, we define a metric of ¡°Stream Coverage Rate
(SCR)¡± as the proportion of stream-based memory accesses in
application¡¯s total accesses:
\begin{equation}
SCR = \frac{Stream\_Accesses}{Total\_Accesses} * 100\%
\end{equation}
Previous works have proposed several stream prefetchers in cache or
memory controller
\cite{Baer95:Prefetch}\cite{Hur06:MemoryPrefetch}\cite{Jouppi90:PrefetchBuffer}\cite{Palacharla94:StreamBuffer}\cite{Smith78:Prefetch}.
However, these proposed techniques are all based on physical address
and little research has focused on impact of virtual address on
stream-based access. Although Dreslinski et al
\cite{Dreslinski07:PrefetchVirtual} have pointed the negative impact
of not accounting for virtual page boundaries in simulation, they
still adapted a non-full system simulator to perform experiments
because of the long period of time to simulate a system in steady
state. Existing research methods have prohibited further
investigations into the impact of OS's virtual memory management on
prefetching. In this case, we have used the HMTT system to reveal
this issue in a real system (Intel Celeron Platform).

\begin{figure}[!t]
\centering
\includegraphics[width=3.5in]{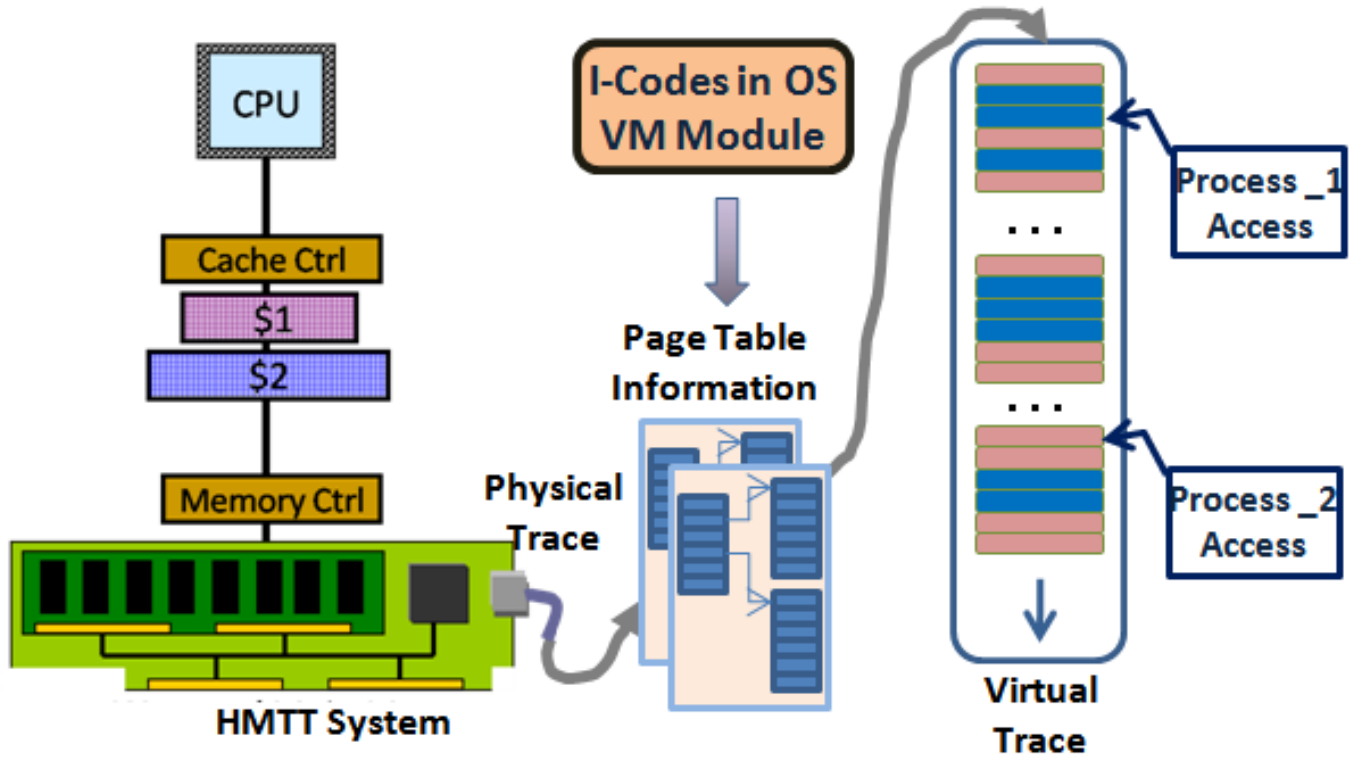}
\caption{A sample of collecting virtual memory trace with the HMTT
system. I-Codes are instrumented into OS's virtual memory management
module to collect page table information. Then the combination of
physical memory trace and page table information can form virtual
memory trace.} \label{fig:collect_virtual_trc}
\vspace{-0.2cm}\end{figure}

Before presenting the case study, we introduce how to use the HMTT
system to collect virtual memory trace. As Figure
\ref{fig:collect_virtual_trc} shown, we insert some I-Codes into
Linux kernel's virtual memory management module to track each page
table entry update. The data is stored in the form of {\bf
\emph{$<$pid, phy\_page, virt\_page, page\_table\_entry\_addr$>$}}
which indicates that a mapping physical page {\bf \emph{phy\_page}}
to virtual page {\bf \emph{virt\_page}} is created for process {\bf
\emph{pid}}, and this mapping information is stored in the location
of {\bf \emph{page\_table\_entry\_addr}}. Thus, given a physical
address, the corresponding process and virtual address can be
retrieved from the page table information. The I-Codes are also
responsible for synchronization with physical memory trace by
referencing the HMTT's configuration space. In this way, we are able
to analyze specified process's virtual memory trace.

\begin{figure}[!t]
\centering
\includegraphics[width=3.5in]{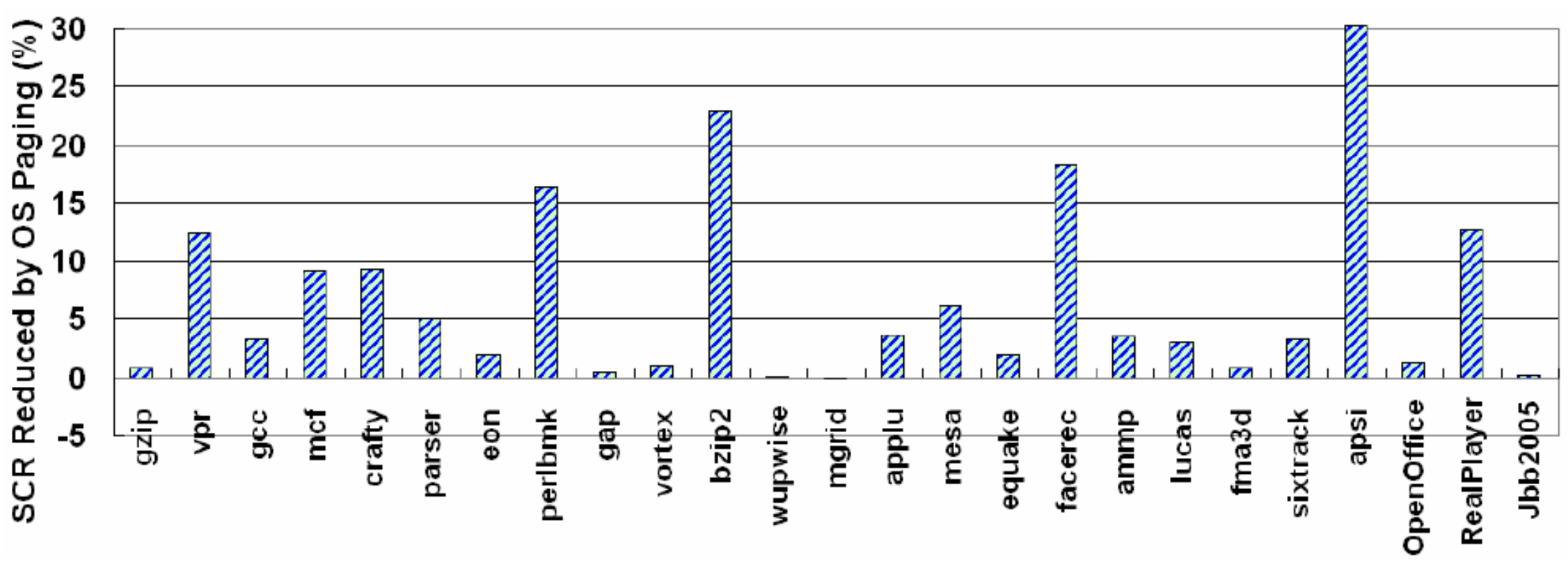}
\caption{The portion of SCR reduced due to OS's virtual memory
management which may map contiguous virtual pages to non-contiguous
physical pages.} \label{fig:SCR_OS} \vspace{-0.2cm}\end{figure}

\begin{figure*}[htbp]
\centering
\includegraphics[width=7in]{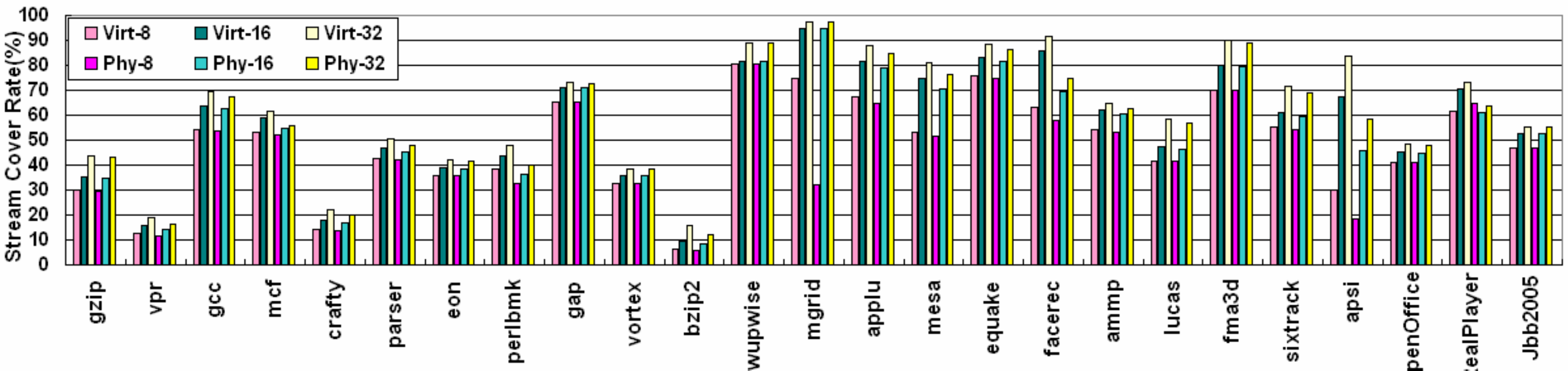}
\caption{"Stream Coverage Rate (SCR)" of various applications under
different detection configurations. For example, the label of
"Virt-8" means detecting SCR among virtual addresses with an 8-entry
detection window, while "Phy-8" means detecting SCR among physical
addresses.} \label{fig:SCR} \vspace{-0.2cm}\end{figure*}

\begin{figure*}[!t]
\centering \vspace{0pt} \subfigure[]{
        \begin{minipage}[t]{0.3\linewidth}
        \centering
        \includegraphics[height=1.5in]{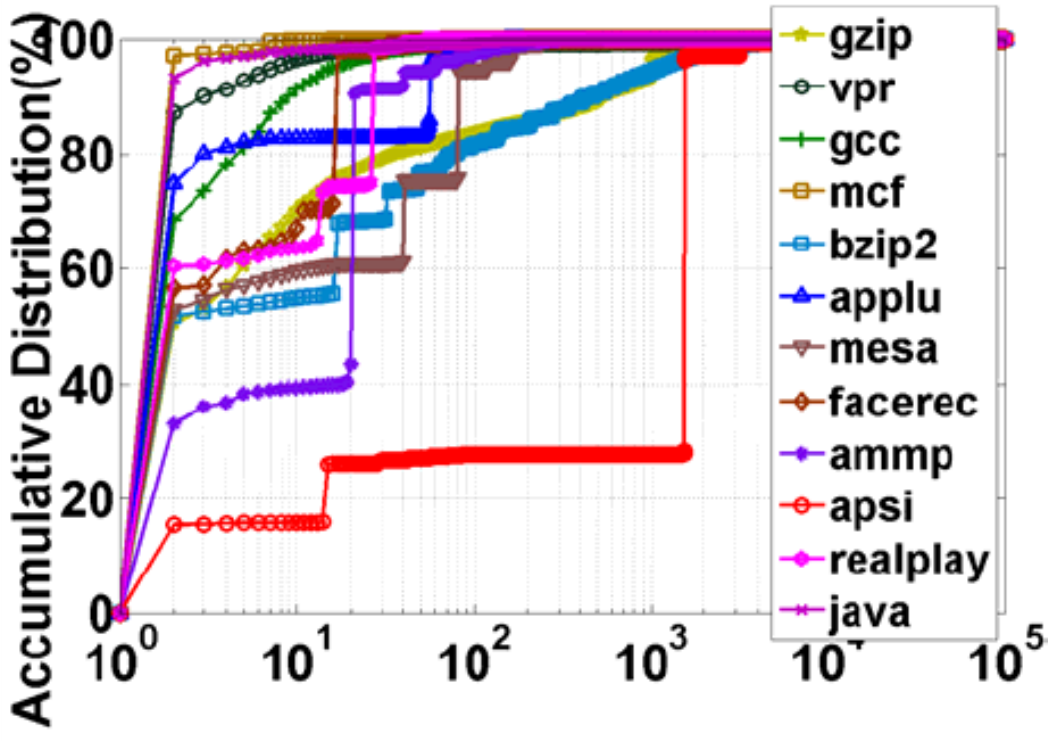}
        \end{minipage}}
\hfill \vspace{0pt} \subfigure[]{
    \begin{minipage}[t]{0.3\linewidth}
    \centering
    \includegraphics[height=1.5in]{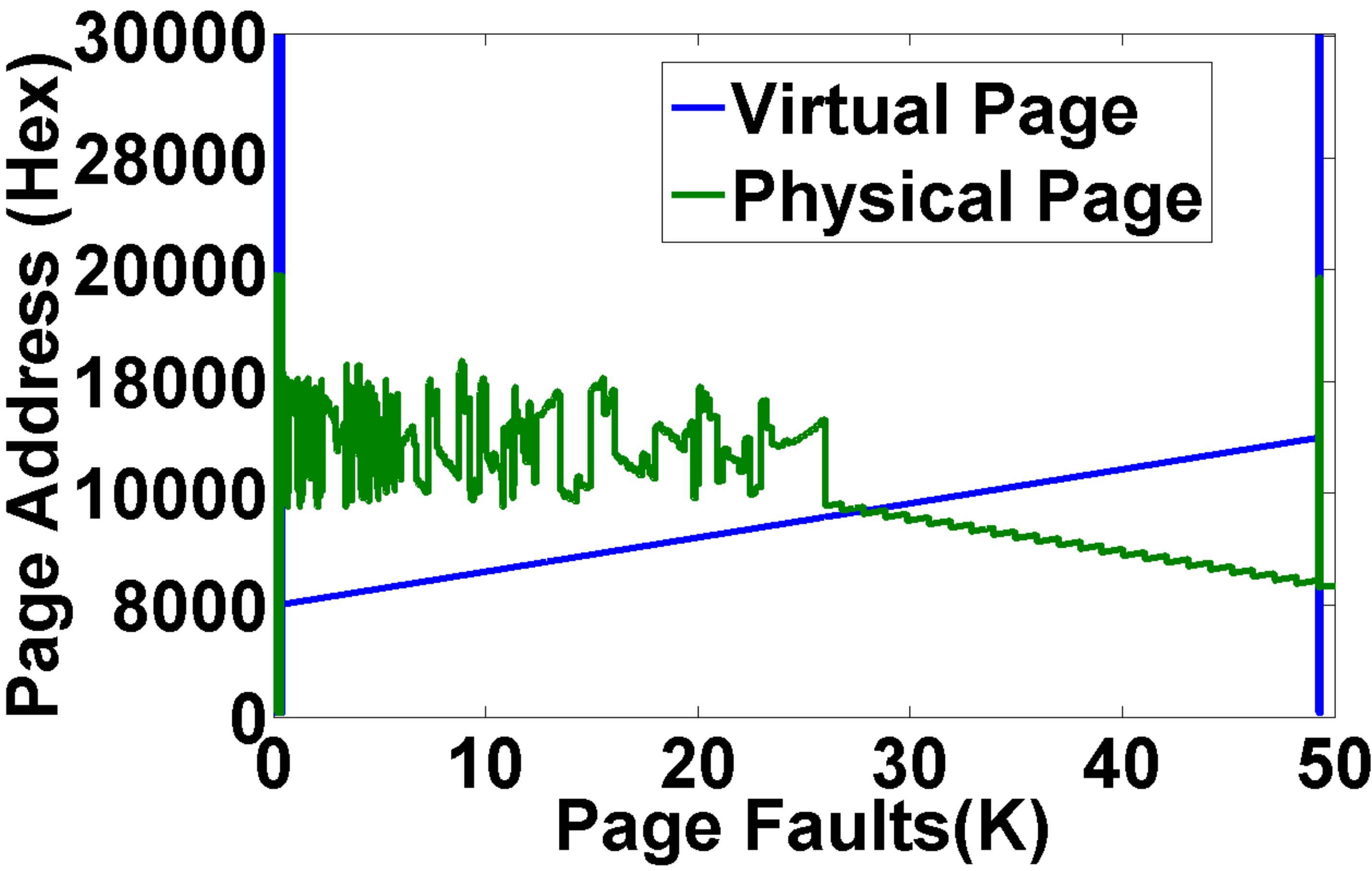}
    \end{minipage}}
\hfill \vspace{0pt} \subfigure[]{
    \begin{minipage}[t]{0.3\linewidth}
    \centering
    \includegraphics[height=1.5in]{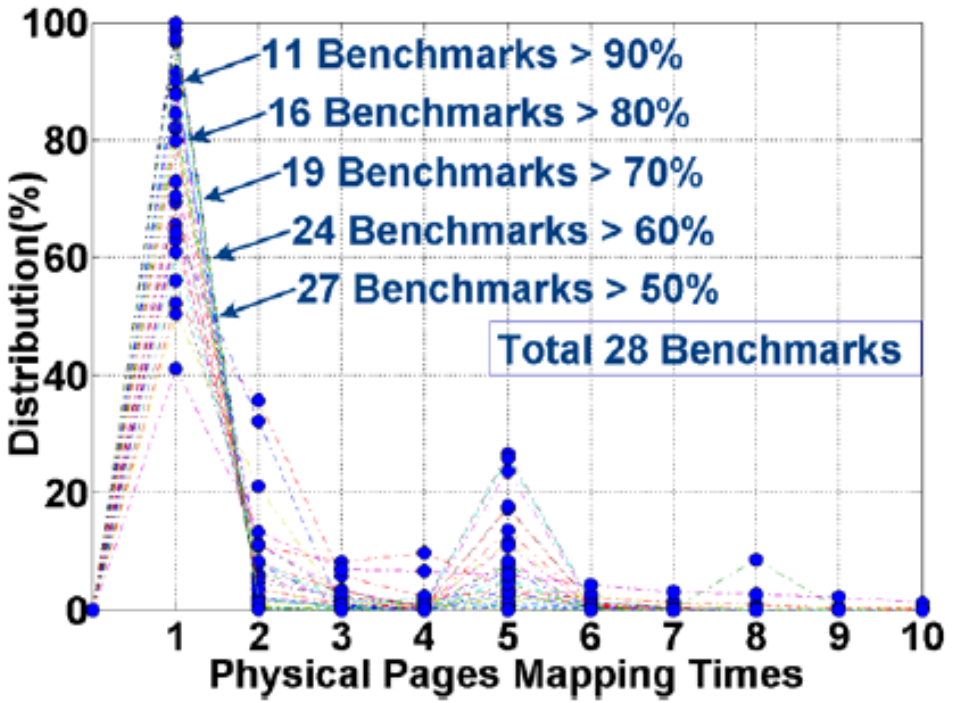}
    \end{minipage}}
\caption{(a) CDF of Stream Stride. (b) The apsi's
virtual-to-physical page mapping information. (c) Distribution of
the virtual-to-physical remapping times during application's
execution lifetime.}\label{fig:SCR_Stride_PF}
\vspace{-0.2cm}\end{figure*}

In order to evaluate application's SCR, we adopt an algorithm
proposed by Mohan et al \cite{Mohan03:Regularity} to detect stream
among L2 cache misses in cache line level.  Figure \ref{fig:SCR}
shows the physical and virtual SCRs detected with different
scan-window sizes. As shown in Figure \ref{fig:SCR}, most
applications' SCRs are more than 40\% under a 32-entry window (The
following studies are based on the 32-entry window). Figure
\ref{fig:SCR_OS} illustrates the portion of SCR reduced due to OS's
virtual memory management. We can see that the OS's influence varies
relying on different applications. Among all 25 applications, the
reduction of 15 applications' SCRs is not significantly, less than
5\%, but there are also 8 applications approaching or exceeding
10\%. As a specifical case, the SCR of "apsi" is reduced by 30.2\%.
We selected several applications to investigate the reason of the
phenomena. Figure \ref{fig:SCR_Stride_PF}(a) shows the CDF of these
applications' stream strides where most applications' strides are
less than 10, within one page. The short strides indicate that most
streams have good spatial locality and also indicate that OS page
mapping may slightly influence the SCRs when streams are within one
physical page. Nevertheless, most strides of the 301.apsi
application are quite large. For example, they is mainly over 64KB
(64B*1000), covering several 4KB-size pages. Figure
\ref{fig:SCR_Stride_PF}(b) illustrates the apsi's
virtual-to-physical page mapping information where virtual pages are
absolutely contiguous but their corresponding physical pages are
non-contiguous.

We can find an interesting observation from Figure
\ref{fig:SCR_Stride_PF}(c) that most physical pages are mapped to
virtual pages only once during application's entire execution
lifetime. This observation implies that either application's working
set has long lifetime or memory capacity is enough so that reusing
physical pages is not required. Thus, to remove the negative impact
of virtual memory management on stream-based access, OS can pre-map
a region where both virtual and physical addresses are linear. For
example, OS can allocate memory within this linear region when
applications call malloc library function. If the region has no
space, OS could determine to either reclaim free space for the
region or allocate in common method.

\subsection{Multicore Impact on Memory Controller}

Programs run on multicore system can concurrently generate memory
access requests. Although L1 cache, TLB etc are usually designed to
be core's private resources, some resources remain sharing by
multiple cores. For example, memory controller is a shared resource
existing in almost all prevalent multicore processors. Thus, memory
controller can receive concurrent memory access requests from
different cores (process/thread). In this case, we will investigate
the impact of multicore on memory controller on the AMD Opteron dual
cores system by the HMTT system. The traces are collected in the
same method as shown in Figure \ref{fig:collect_virtual_trc}, which
is depicted in last section .

\begin{figure*}[!t]
\centering \vspace{0pt} \subfigure[]{
    \begin{minipage}[t]{0.29\linewidth}
    \centering
    \includegraphics[height=1.13in]{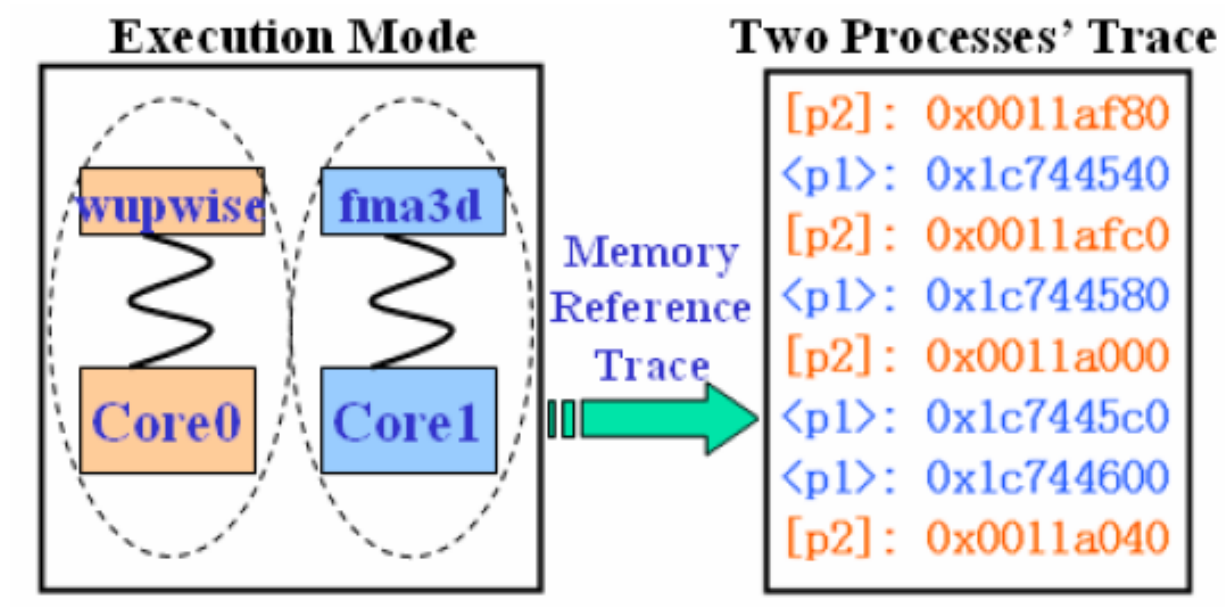}
    \end{minipage}}
\hfill \vspace{0pt} \subfigure[]{
        \begin{minipage}[t]{0.22\linewidth}
        \centering
        \includegraphics[height=1.13in]{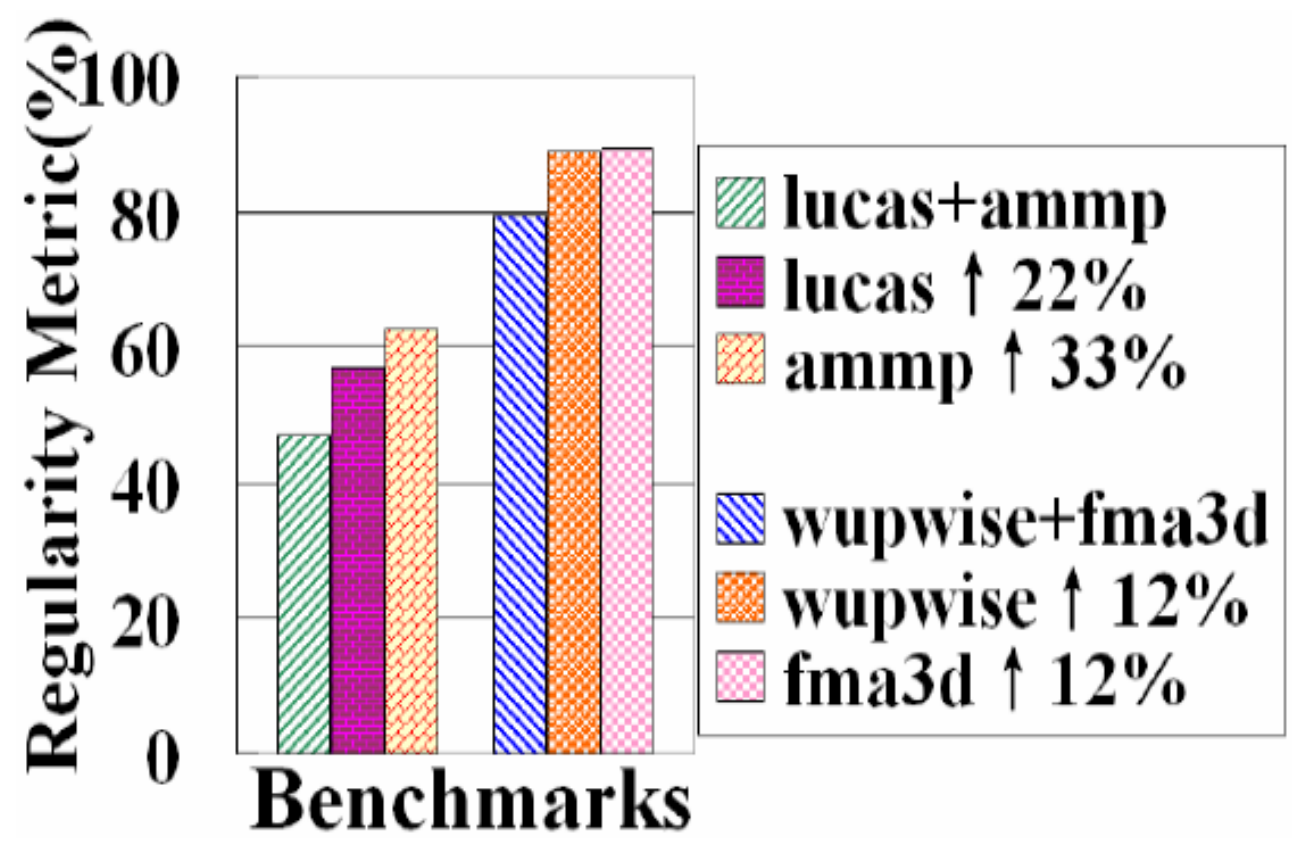}
        \end{minipage}}
\hfill \vspace{0pt} \subfigure[]{
    \begin{minipage}[t]{0.2\linewidth}
    \centering
    \includegraphics[height=1.13in]{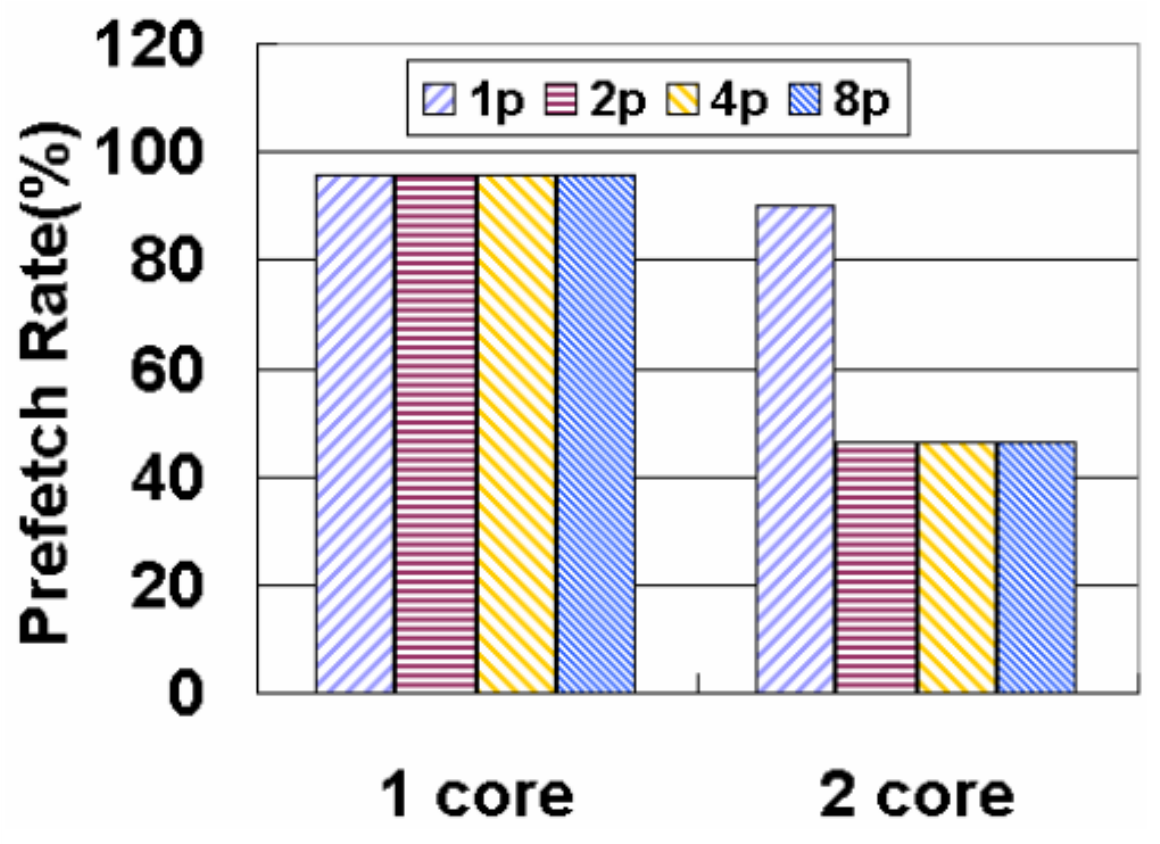}
    \end{minipage}}
\hfill \vspace{0pt} \subfigure[]{
    \begin{minipage}[t]{0.2\linewidth}
    \centering
    \includegraphics[height=1.13in]{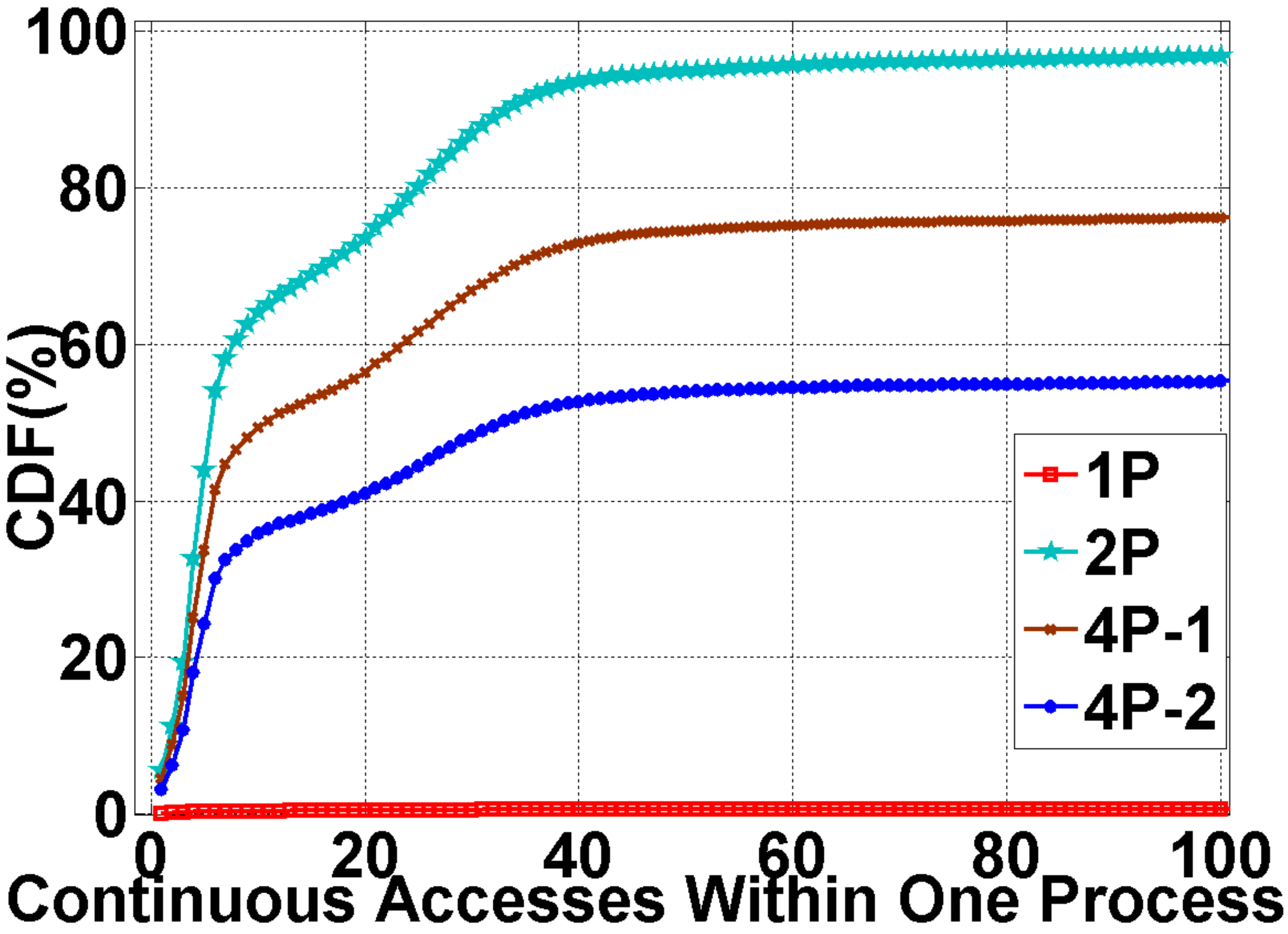}
    \end{minipage}}
\caption{(a) A slice of memory trace presents the phenomenon of
interleaved memory accesses between two processes. (b) The memory
controller can detect better regularity metric (SCR) if it is aware
of processes. (c) Perform micro-benchmarks to investigate the effect
of AMD's sequential prefetcher in memory controller. (d) CDF of
continuous memory accesses of one process before interfered by
another process. }\label{fig:Multicore_MC}
\vspace{-0.2cm}\end{figure*}

Figure \ref{fig:Multicore_MC}(a) illustrates a slice of memory trace
of two concurrent processes, i.e. wupwise and fma3d, where the
phenomenon of interleaved memory access is intuitively obvious. We
detect "Stream Coverage Rate (SCR)" of the two-process mixed trace
and find that the SCR can is about 46.9\% and 79.3\% for
$<$lucas+ammp$>$ and $<$wupwise+fma3d$>$ respectively. Because the
memory traces collected by the HMTT system contain process
information, we are able to detect SCR of individual process's
memory accesses. In this way, we evaluate the potential of
process-aware detection policy at memory controller level. Figure
\ref{fig:Multicore_MC}(b) shows that the SCRs of lucas and ammp
increase to 56.9\% and 62.6\% respectively, as are the SCRs of
wupwise and fma3d.

Further, we adopt micro-benchmarks to investigate the effect of
AMD's sequential prefetcher in memory controller. We use performance
counters to collect statistic of two events, i.e. {\bf
\emph{DRAM\_ACCESSES}} which indicates the number of memory requests
issued by memory controller and {\bf \emph{DATA\_PREFETCHES}} which
indicates the number of prefetching requests issued by sequential
prefetcher. Here, we define a metric of {\bf \emph{Prefetch\_Rate}}
as the proportion of prefetching requests in total memory accesses:
\begin{equation}
Prefetch\_Rate = \frac{DATA\_PREFETCHES}{DRAM\_ACCESSES} * 100\%
\end{equation}
It should be noted that the sequential prefetcher can issue
prefetching requests only after it detects three contiguous
accesses. Thus, the {\bf \emph{Prefetch\_Rate}} also implies the
"Stream Coverage Rate (SCR)" detected by the prefetcher.

\begin{figure}[!t]
\centering
\includegraphics[width=3in]{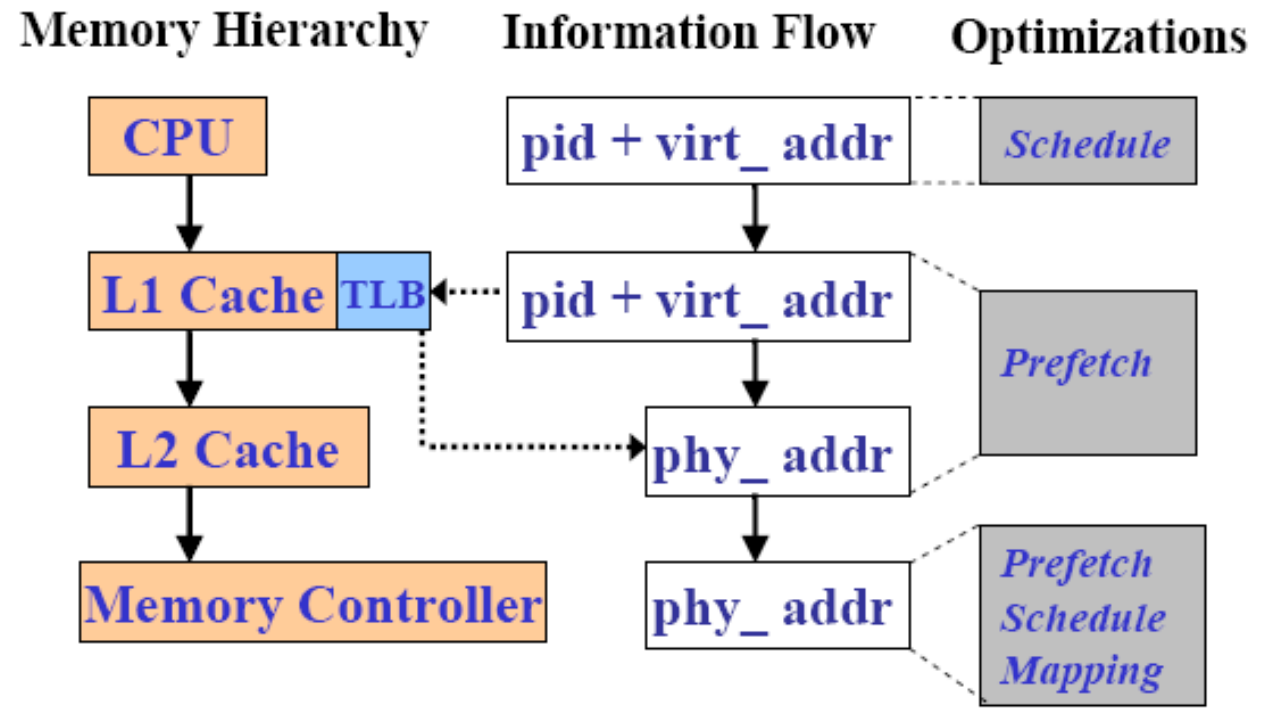}
\caption{Memory Access Information Flow. In traditional memory
hierarchy, access information is continually reduced from high
memory hierarchy level to low level.} \label{fig:mem_access_flow}
\vspace{-0.2cm}\end{figure}

We run a micro-benchmark to sequentially read a linear 256MB
physical memory region. Since our experimental AMD Opteron processor
has two cores, we disable one core to emulate one-core environment
by Linux's tool. Intuitively, an ideal sequential prefetcher should
achieve a {\bf \emph{Prefetch\_Rate}} of near 100\%. As shown in
Figure \ref{fig:Multicore_MC}(c), this ideal case exists in one-core
environment no matter how many processes run concurrently. However,
in the two-core environment, the {\bf \emph{Prefetch\_Rate}} of
one-precess case is still over 90\% but it sharply decrease to 46\%
when running two or more processes concurrently. We further
investigate the phenomenon by analyzing memory trace. Figure
\ref{fig:Multicore_MC}(d) shows the CDF of one process' continuous
memory accesses before interfered by another process. We can find
that when two processes run concurrently, in most cases (over 95\%)
memory controller can only handle less than 40 memory access
requests of one process and then will be interrupted to handle
another process' requests. For the case of running one process,
memory controller can handle over 1000 memory access requests of the
process and then be interrupted to handle other requests. These
experiments reveal that the interference of memory accesses from
multiple cores (i.e., processes/threads) is serious.

In a word, although prefetcher has been integrated into memory
controller for optimizing memory system performance, it cannot
produce an expected effect if not consider the multicore's impact.
Usually, optimization requires request information. Figure
\ref{fig:mem_access_flow} shows a traditional memory access
information flow in a common memory hierarchy. We can find that
memory access information is continually reduced when a request
passes from high memory hierarchy level to low level. For example,
after TLB's address translation, L2 cache and memory controller can
only obtain physical address information. So if more information
(e.g., core-id, virtual address) could be passed through the memory
hierarchy, those optimization techniques for low-level hierarchies
(L2/L3 cache and memory controller) should gain better effect.

\subsection{Characterization of DMA Memory Reference}

I/O accesses are essential on modern computer systems, whenever we
load binary files from disks to memory or download files from
network. DMA technique is used to release processor from I/O
process, which provides special channels for CPU and I/O devices to
exchange I/O data. However, as the throughput of the I/O devices
grows rapidly, memory data moving operations have become critical
for DMA scheme, which becomes a performance bottleneck for I/O
operations. In this case, we will investigate the characterization
of DMA memory reference.

\begin{figure}[!t]
\centering
\includegraphics[width=3.5in]{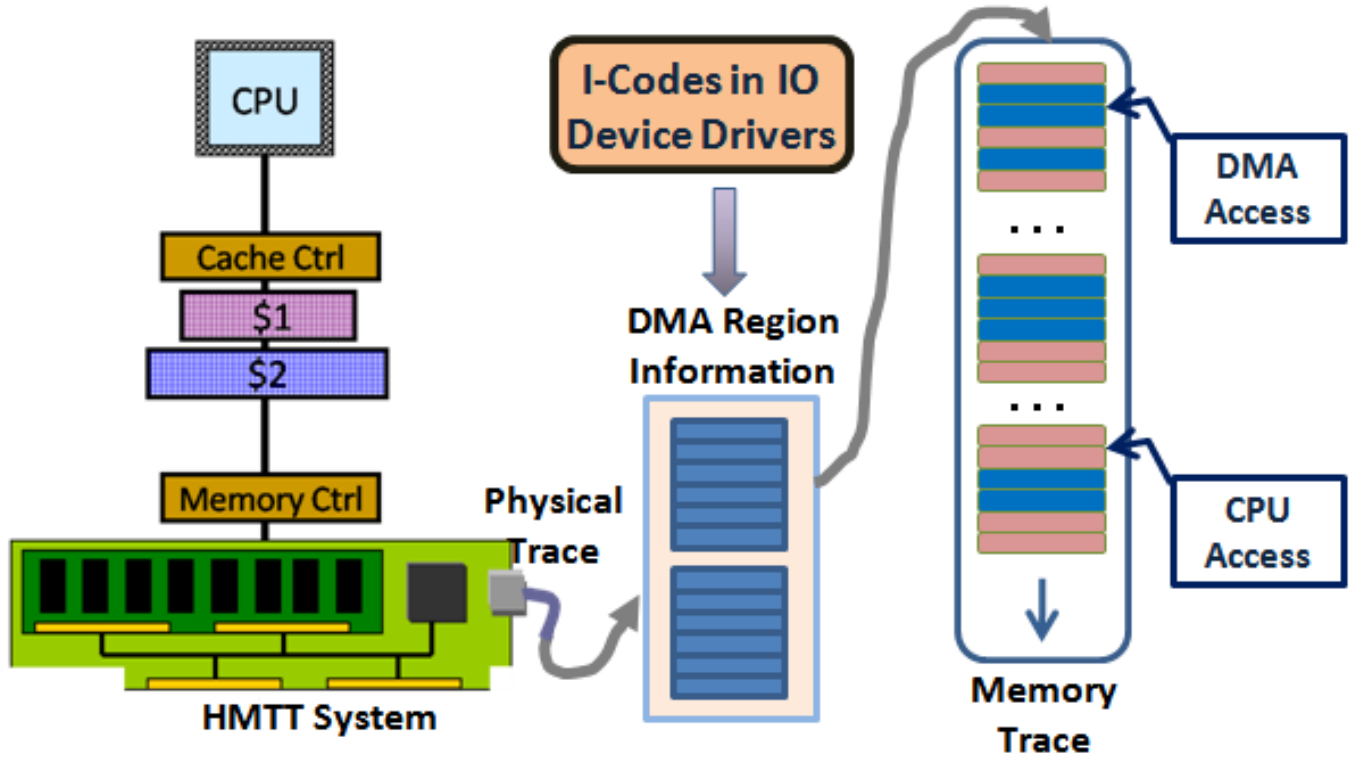}
\caption{A sample of distinguishing CPU and DMA memory trace with
the HMTT system. I-Codes are instrumented into I/O device drivers to
collect DMA request information, e.g., DMA buffer region, the time
when allocate and release DMA buffer. Holding the information, we
can identify those memory accesses falling into both space region
and time interval as DMA memory accesses.}
\label{fig:collect_dma_trc}
\end{figure}

First, we introduce how to collect DMA memory reference trace. To
distinguish a memory reference issued by DMA engine or processor, we
have inserted I-Codes into the device drivers of hard disk
controller and network interface card (NIC) on Linux platform.
Figure \ref{fig:collect_dma_trc} illustrates the memory trace
collection framework. When the modified drivers allocate and release
DMA buffers, the I-Codes record start address, size and owner
information of a DMA buffer. Meanwhile, they send synchronization
tags to the HMTT system's configuration space. When the HMTT system
receives synchronization tags, it injects tags (DMA\_BEGIN\_TAG or
DMA\_END\_TAG) into physical memory trace to indicate that those
memory references between the two tags and within the DMA buffer¡¯s
address region are DMA memory references initiated by DMA engine.
The status information of DMA requests, such as start address, size
and owner, is stored in a reserved kernel buffer and is dumped into
a file after memory trace collection is completed. Thus, there is no
interference of additional I/O access. In this way, we can
differentiate DMA memory reference from processor memory reference
by merging physical memory trace and status information of DMA
requests. In this case, we run all the benchmarks on the AMD server
machine and use the HMTT system to collect memory reference traces
of three real applications (file-copy, TPC-H, and SPECweb2005).

Table \ref{Table:percentage_of_memory_request_type} shows the
percentages of DMA memory references in various benchmarks. In Table
\ref{Table:percentage_of_memory_request_type} we can see that the
file-copy benchmark has nearly the same percentage of DMA read
references (15.4\%) and DMA write references (15.6\%), and the sum
of two kinds of DMA memory references is about 31\%. For TPC-H
benchmark, the percentage of all DMA memory references is about
20\%.  The percentage of DMA write references (19.9\%) is about 200
times of that of DMA read references (0.1\%) because the dominant
I/O operations in TPC-H is transferring data from disk to memory
(i.e., DMA write request). For SPECweb2005, the percentage of DMA
memory references is only 1.0\%. Because the size of network I/O
requests (including a number of DMA memory references) is quite
small, processor is busy with handling interrupts.

\begin{table}[!t]\small
\centering \caption{Percentage of Memory Reference Type}
\label{Table:percentage_of_memory_request_type}
\begin{tabular}{|l||r|r|r|}
  \hline
                 & File Copy     &  TPC-H &  SPECweb2005    \\
                 \hline
  CPU Read       &   45\%       &   60\%  &    75\%         \\
                 \hline
  CPU Write      &   24\%       &   20\%  &    24\%         \\
                 \hline
  DMA Read       & 15.4\%       & 0.1\%  &   0.76\%        \\
                 \hline
  DMA Write      & 15.6\%       & 19.9\%  &   0.23\%        \\
                 \hline
\end{tabular}
\end{table}

\begin{figure}[!t]
\centering
    \includegraphics[width=3.5in]{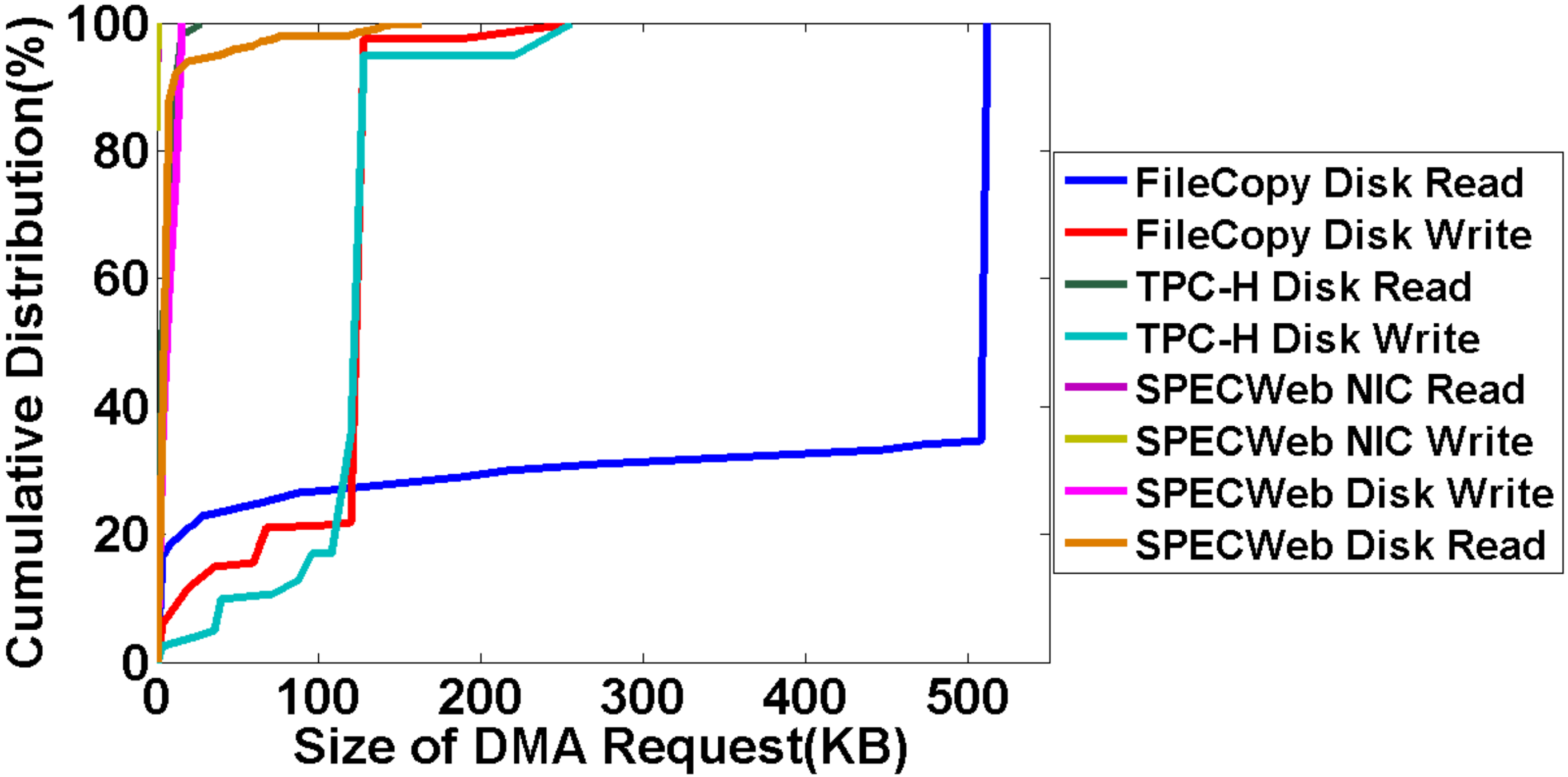}\\
    \caption{Cumulative Distribution of DMA Request Size}\label{Figure:dma_trace_character}
\vspace{-0.2cm}\end{figure}

Table \ref{Table:average_size_of_DMA} and Figure
\ref{Figure:dma_trace_character} depict the average size of DMA
requests and the cumulative distributions of the size of DMA
requests for three benchmarks respectively (one DMA request includes
a number of DMA memory references with 64 bytes). For file-copy and
TPC-H benchmarks, all DMA write requests are less than 256KB and the
percentage of those requests with the size of 128KB is about 76\%.
The average sizes of DMA write requests are about 110KB and 121KB
for file-copy and TPC-H respectively. For SPECweb2005, the size of
all DMA requests issued by NIC are smaller than 1.5KB because the
maximum transmission unit (MTU) of Gigabit Ethernet frame is only
1518 bytes. The size of DMA requests issued by IDE controller for
SPECweb2005 is also very small, an average of about 10KB.

\begin{table}[!t]\small
\centering \caption{Average Size of Various Types of DMA Requests}
\label{Table:average_size_of_DMA}
\begin{tabular}{|r||l|r|r|}
  \hline
                 & Request Type   &  \% &  Avg Size    \\
                 \hline
  File           & Disk DMA Read   &   49.9    &    393KB         \\
  Copy           & Disk DMA Write  &   50.1    &    110KB         \\
                 \hline
  TPC-H          & Disk DMA Read   &    0.5    &     19KB         \\
                 & Disk DMA Write  &   99.5    &    121KB         \\
                 \hline
                 & Disk DMA Read   &   24.4    &     10KB         \\
  SPECweb        & Disk DMA Write  &    1.7    &      7KB         \\
  2005           & NIC DMA Read   &     52    &    0.3KB         \\
                 & NIC DMA Write  &   21.9    &    0.16KB         \\
                 \hline
\end{tabular}
\end{table}

It should be noted that some studies have focused on reducing the
overhead of additional memory copy operations for I/O data transfer,
such as Direct Cache Access (DCA) \cite{DCA:05Huggahalli}. However,
their study focuses on network traffics and shows that the DCA
scheme has poor performance for applications that have intensive
disk I/O traffics (e.g. TPC-C). Our evaluations have shown this is
because sizes of disk DMA requests (100+KB) are larger than those of
network ($<$1KB). Therefore, disk I/O data can cause serious cache
interference. We will further analyze and optimize I/O memory access
in our future work.

\subsection{Summary of Case Studies}
In this section, we have present three case studies to demonstrate
the widespread and effective use of the HMTT system. It should be
noted that although we insert these I-Codes into OS modules and
device drivers manually in the three case studies, we are enhancing
the HMTT system by integrating binary instrumentation into it. In
this way, the HMTT system is able to collect information from binary
files. Overall, the case studies have shown the HMTT System, which
adopts the hybrid hardware/software tracing mechanism, is a feasible
and convincing memory trace monitoring system.

\section{Related Work}

There are several areas of effort related to memory trace
monitoring: software simulators, binary instrumentation, hardware
counters, hardware monitors and hardware emulators.

{\bf Software simulators:} Most memory performance and power
researches are based on simulators. They utilize cycle-accurate
simulators to generate memory trace and then feed trace to
tracedriven memory simulators (e.g. DRAMSim \cite{Wang05:DRAMSim},
MEMsim \cite{Rajamani00:MEMSim}). SimpleScalar \cite{SimpleScalar}
is a popular user-level simulator, but it can not run operating
system for analysis of full system behaviors. Several full system
simulators (such as SimOS \cite{Rosenblum95:SimOS}, Simics
\cite{Magnusson02:Simics}, M5 \cite{M5}, BOCHS \cite{BOCHS} and QEMU
\cite{Bellard05:QEMU}), which can boot commercial operating systems,
are commonly used in research when deal with OS-intensive
applications. However, software simulators usually have speed and
scalability limitations. As the computer architectures become more
and more sophisticated, more detail simulation models are need,
which may lead to a slowdown of 1000X$\sim$10000X
\cite{Barroso99:EvaluationArch}. Moreover, simulation with complex
multicore and multi-threaded applications may incur inaccuracies and
could lead to misleading conclusions \cite{Nanda00:MemorIES}.

{\bf Binary instrumentation:} Many binary instrument tools (e.g.
O-Profile \cite{OProfile}, ATOM \cite{Srivastava94:ATOM}, DyninstAPI
\cite{DyninstAPI}, Pin \cite{PIN-tool}, Valgrind \cite{Valgrind},
Nirvana \cite{Bhansali06:inst-Tracing} etc.) are popularly utilized
to profile applications. They are able to obtain applications¡¯
virtual access trace even without source codes. Nevertheless, few of
them can provide full system memory trace because instrumenting
kernels is very tricky. PinOS \cite{Bungale07:PinOS} is an extension
of the Pin \cite{PIN-tool} dynamic instrumentation framework for
full-system instrumentation. It is built on top of the Xen
\cite{Barham07:Xen} virtual machine monitor with Intel VT
\cite{Neiger06:IntelVT} technology and, is able to instrument both
kernel and user-level code. However, PinOS can only run on IA-32 in
uni-processor mode. Moreover, binary instrumentation method usually
slows down target programs¡¯ execution, incurring time distortion
and memory access interference.

{\bf Hardware counters:} Hardware counters are able to provide
accurate events statistic (e.g. Cache Miss, TLB Miss, etc.).
Itanium2 \cite{Itanium2} is even able to collect trace via sampling.
The approach of hardware counters is fast, low overhead, but they
can not track complete and detailed memory reference trace.

{\bf Hardware monitors:} Various Hardware monitors, divided into two
classes, are able to monitor memory trace online. One class is pure
trace collectors, and another is online cache emulators. BACH
\cite{Flanagan92:BACH}\cite{Grimsrud93:BACH} is a trace collector.
It utilizes a logic analyzer to interface with host system and to
buffer the collected traces. When the buffer is full, the host
system is halted by an interrupt and the trace is moved out. Then,
the host system continues to execute programs. BACH is able to
collect traces from long workload runs. However, this halting
mechanism may alter original behavior of programs. The
hardware-based online cache emulation tools (such as MemorIES
\cite{Nanda00:MemorIES}, PHA\$E \cite{Chalainanont03:PHASE}, RACFCS
\cite{Youn97:RACFCS}, ACE \cite{Hong06:ACE}, and HACS
\cite{Watson02:HACS}) are very fast and have low distortion and no
slowdown. Logic analyzer is also a powerful tool for capturing
signals (including DRAM signals) and can be very useful for hardware
testing and debugging.

However, these hardware monitors have several disadvantages: (1)
they (except BACH) are not able to dump full mass trace but only
produce short traces due to small local memories; (2) there is a
semantic gap problem for hardware monitors because they can only
collect physical address; (3) they depend on proprietary interfaces,
for example, MemorIES relies on the IBM¡¯s 6xx bus, BACH, PHA\$E,
ACE, HACS etc. adopt logic analyzer which is quite expensive. RACFCS
use a latch board that directly connects to output pins of specified
CPUs. So they have poor portability.

{\bf Hardware emulators:} Several hardware emulators are thorough
FPGA-based systems which utilize a number of FPGAs to construct
uni-processor/multi-processor research platforms to accelerate
research. For example, RPM \cite{Barroso95:RPM} emulates the entire
target system within its emulator hardware. Intel proposed an
FPGA-based Pentium system \cite{Lu07:FPGA-Pentium} which is an
original Socket-7 based desktop processor system with typical
hardware peripherals running modern operating systems. RAMP
\cite{RAMP} is also a new scheme for architecture research. Although
they do not produce any memory traces currently, they are capable of
tracking full system trace. But they can only emulate a simplified
and slow system with relative fast I/O, which enlarges the
¡°CPU-memory / memory-disk¡± gaps that may be bottlenecks in real
systems.

\section{Conclusion}
In this paper we propose a hybrid hardware/software mechanism which
is able to collect memory reference trace as well as semantic
information. Based on this mechanism, we have designed and
implemented a prototype system called HMTT (Hybrid Memory Trace
Tool) which adopts a DIMM-snooping mechanism to snoop on memory bus
and a software-controlled trace injection mechanism capable of
injecting semantic information into normal memory trace.
Comprehensive validations show that the HMTT system is a feasible
and convincing memory trace monitoring system. Several case studies
show that it is also effective and widespread. Thus, the HMTT system
demonstrates that the hybrid tracing mechanism can leverage both
hardware's (e.g., no distortion or pollution) and software's
advantages (e.g., flexibility and more information). Moreover, this
hybrid mechanism can be adopted by other tracing systems.


%


\ifCLASSOPTIONcompsoc
\else
\fi


\ifCLASSOPTIONcaptionsoff
  \newpage
\fi



\bibliographystyle{abbrv}
\bibliography{./bib/tex}
%
%
%

\appendices
\section{HMTT's Verification and Evaluation}
\subsection{Verification}
The HMTT system is verified in four steps:

1) As a basic verification, we have checked the physical address
trace tracked by the monitoring board (MTB) with micro benchmarks
which generate sequential reads, sequential writes, sequential
read-after-writes and random reads in various unites from cache line
to page size. The test results show that there are no incorrect
physical addresses.

\begin{figure}[htbp]
\centering
\includegraphics[width=3.5in]{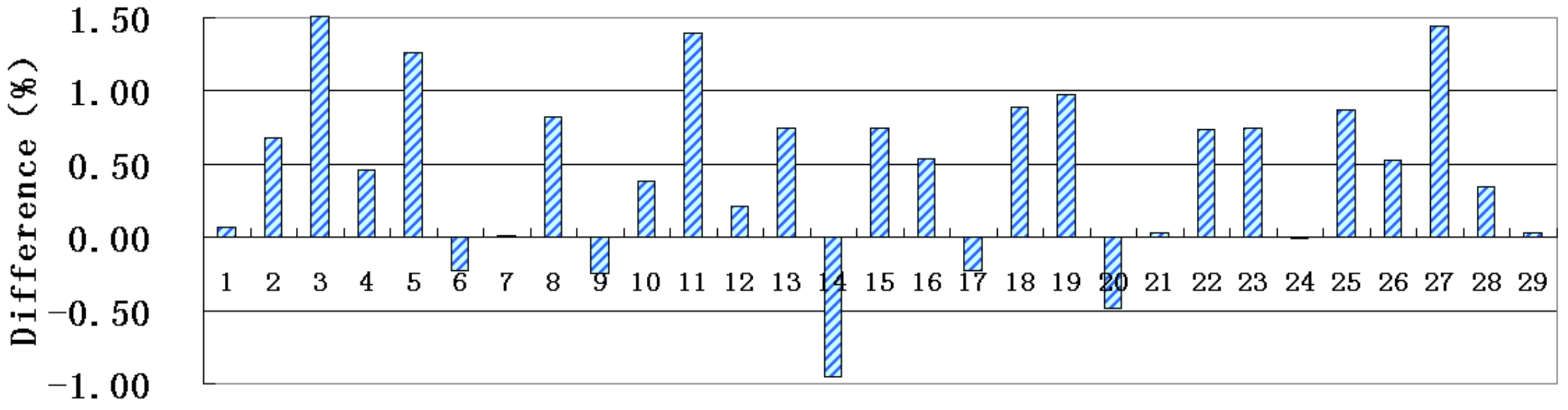}
\caption{ A comparison with Performance Counter. While running 29
programs of SPECCPU2006 \cite{SPEC:CPU2006}, we compare the numbers
of memory references collected by HMTT and the numbers of
DRAM\_ACCESS events collected by O-Profile \cite{OProfile}.}
\label{fig:hw_cnt_diff} \vspace{-0.2cm}\end{figure}

2) A comparison with performance counter (use O-Profile
\cite{OProfile} with DRAM\_ACCESS event) is illustrated in Figure
\ref{fig:hw_cnt_diff}. Note that the axis represent 29 programs of
SPECCPU2006 \cite{SPEC:CPU2006}. Through the figure, differences of
memory access numbers acquired by HMTT and performance counter
respectively are mostly less than 1\%, mainly incurred in
initialization and finalization phases.

\begin{figure}[htbp]
\centering
\includegraphics[width=3.5in]{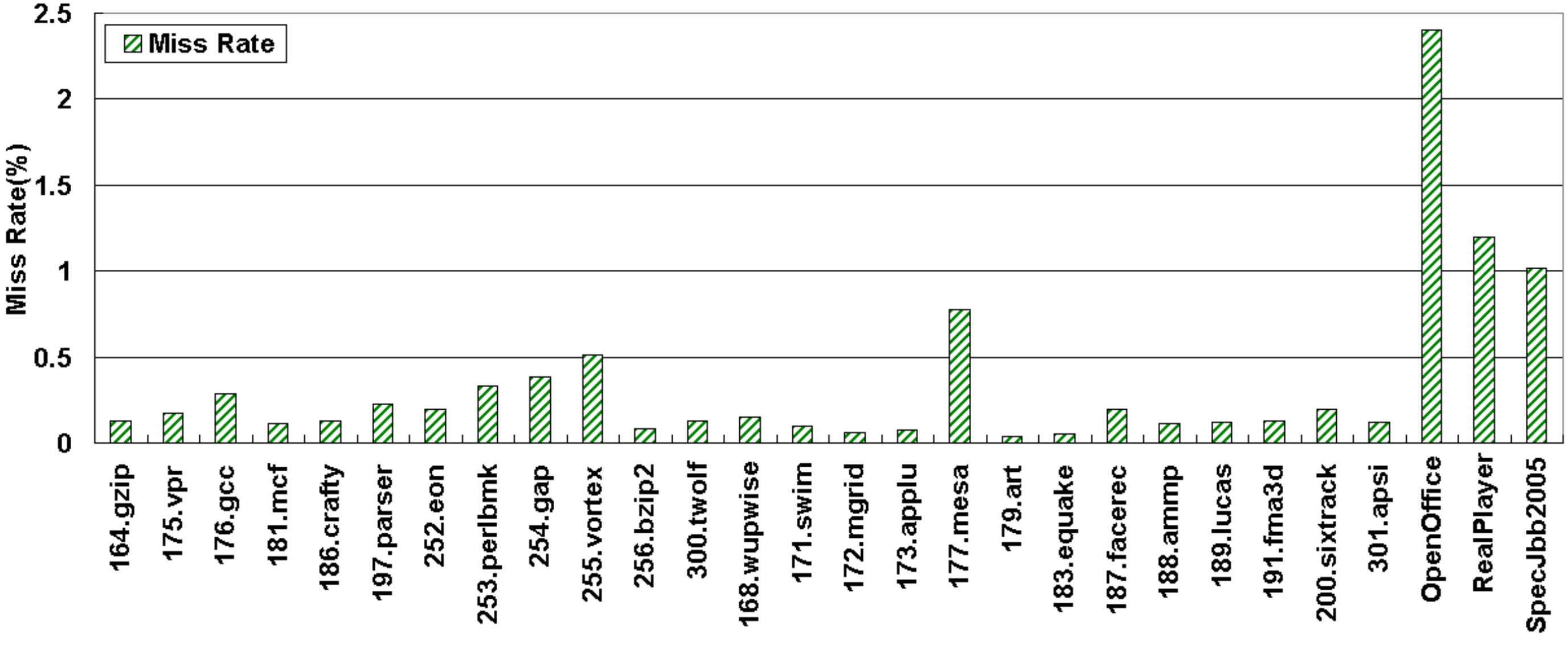}
\caption{Various Applications' Miss Rates. Here, "Miss Rate"
indicate the portion of those physical addresses cannot be
translated to virtual addresses due to unmapped I/O memory
references. } \label{fig:miss_rate} \vspace{-0.2cm}\end{figure}

\begin{figure}[!t]
\centering
\includegraphics[width=3.5in]{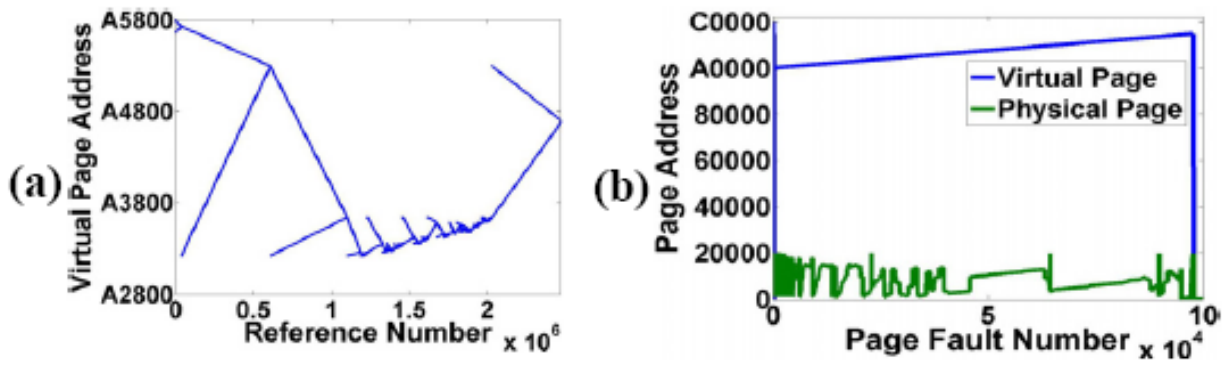}
\caption{An example of QuickSort program with an input of 10M
integers. (a) QuickSort's virtual memory reference pattern; (b)
QuickSort's page table information -- virtual-to-physical page
mapping.} \label{fig:quicksort} \vspace{-0.2cm}\end{figure}

3) The following two steps are to verify software parts of the HMTT
system. Here, we present a case of virtual address trace
verifications. To obtain virtual address trace, we adopt an
assistant kernel module to collect page table information. We have
replayed virtual memory trace to verify if physical addresses and
virtual address are corresponding. Figure \ref{fig:quicksort} shows
an example of quicksort¡¯s virtual memory reference trace with an
input of 100,000,000 integers. Figure \ref{fig:quicksort}(b) shows
the virtual address space and its corresponding physical address
space of quicksort¡¯s data segment collected by the kernel module.
The virtual address region is linear but the physical address region
is discrete. Figure \ref{fig:quicksort}(a) shows a piece of virtual
memory trace, which presents the exact reference pattern of
quicksort. Moreover, the address space region (0xA2800$\sim$0xA5800)
also belongs to the virtual address space of data segment
(0xA0000$\sim$0xC0000) (Figure \ref{fig:quicksort}(b)).

4) Figure \ref{fig:miss_rate} shows miss rates which indicate the
portion of those physical addresses that cannot be translated to
virtual addresses. These "misses" are generated due to some I/O
operations which are performed without page mapping. As Figure
\ref{fig:miss_rate} shown, the miss portions of SPECCPU are nearly
all less than 0.5\%, while those applications with more I/O accesses
have miss portions of over 1\%. Note that we also introduce other
I-Codes (see figure \ref{fig:hmtt_design}) to further distinguish
I/O memory reference, which will be discussed later.

The above verification works show that the HMTT system is a feasible
and convincing memory tracing system.

\subsection{Evaluations}
The trace bandwidth is a crucial issue for the HMTT system. We first
adopt a mathematical method to analyze the trace bandwidth issue. We
let {\bf \emph{BW}} denote trace bandwidth, {\bf \emph{cmdfrq}}
denote the frequency of DDR read/write commands \footnote{Note that
trace is only generated on read/write commands.}, {\bf
\emph{tracenum}} denote the number of trace generated upon each DDR
command and {\bf \emph{bitwidth}} denote the bitwidth of each trace.
Then we can calculate the trace bandwidth in the following equation:
\begin{equation}\label{eq:BW}
BW = cmdfrq~*~tracenum~*~bitwidth
\end{equation}
Next, we will present how to determine the values of the three
parameters. According to the timing diagram from JEDEC specification
\cite{DDR2JEDEC}, we can find that the maximal frequency of DDR
read/write command is dependent on the parameter of CAS-CAS delay
time (tCCD). On the other hand, read and write accesses to the DDR2
SDRAM are burst oriented, which means that accesses start at a
selected location and continue for a burst length (BL) of four or
eight in a certain sequence. Thus, we can get that:
\begin{equation}
cmdfrq = \frac{FREQ_{mem}}{max\{2*tCCD, BL\}}
\end{equation}
In practise, {\bf \emph{BL}} (Burst Length) is larger than {\bf
\emph{2*tCCD}}\footnote{For DDR/DDR2, usually BL is equal to 4 or 8
and the tCCD is equal to 2. For DDR3, tCCD is equal to be 4.}. So we
can calculate the data transferred on memory bus upon each
read/write command as {\bf \emph{BL*$WIDTH_{membus}$}}. Because most
memory controllers handle memory requests to read/write whole cache
line which can be identified as one memory trace, the parameter of
{\bf \emph{tracenum}} denoting the number of trace generated upon
each read/write command can be calculated as follows:
\begin{equation}
tracenum = \frac{BL*WIDTH_{membus}}{SIZE_{cacheline}}
\end{equation}
Since {\bf \emph{BL $\ge$ 2*tCCD}}, Equation (\ref{eq:BW}) is
rewritten:
\begin{equation}
BW = \frac{FREQ_{mem}~*~WIDTH_{membus}~*~bitwidth}{SIZE_{cacheline}}
\end{equation}
For instance, when we set trace {\bf \emph{bitwidth}} to be 40 bits,
the trace bandwidth for a dual-channel DDR2-400 machine with 128-bit
memory bus and 64-byte cache line can be calculated as follows:
\begin{equation}
BW_{ddr2-400}=\frac{400MHz*128bits*40bits}{64bytes}=4Gb/s
\end{equation}
It should be noted that this is the peak trace bandwidth. In
practise, because applications have occasional burst memory access
phases, we find that it is sufficient to adopt a 16K-entry FIFO to
buffer traces and three Gigabit Ethernet interfaces in the HMTT
system to send memory trace.


In  hundreds of experiments, we have verified that bandwidths of
3Gb/s and 1Gb/s are sufficient for DDR2-400 and DDR-200
respectively. Table \ref{Tab:Trace_data} illustrates trace
generation bandwidth of various applications on two DDR-200MHz
machine machines (SPECCPU2000, desktop and server applications are
on an Intel Celeron machine and SPECCPU2006 is on an AMD Opteron
machine). The bandwidth varies from 5.7MB/s (45.6Mbps) to 72.9MB/s
(583.2Mbps) on Intel platform, and from 0.1MB/s (0.8Mbps) to
106.8MB/s (854.4Mbps) on AMD platform. This indicates that a
bandwidth of 1Gb/s is sufficient for the HMTT system to capture all
applications¡¯ traces on Intel platform and most applications¡¯
traces on AMD platform. However, the high frequency of DDR2/DDR3
memory and prevalent multi-channel memory technology increase trace
data generation bandwidth. Therefore, the HMTT system supports three
Gigabit Ethernet interfaces currently and will adopt PCI-E inerface
to provide a bandwidth of 10Gb/s to overcome the bandwidth problem.

\begin{table}
\centering
  \caption{Trace Generation Bandwidth}\label{Tab:Trace_data}
\begin{tabular}{|c|l|c||c|l|c|}
  \hline
        & Appication  &   BW & &  Appication & BW\\
        &              & (MB/s)    & &      & (MB/s)  \\
 \hline
 \hline
 S & 164.gzip    & 33.9  & S   & 168.wupwise & 24.6   \\
 P & 165.vpr     & 44.9  & P   & 171.swim    & 65.8 \\
 E & 176.gcc     & 44.5  & E  & 172.mgrid   & 48.0   \\
 C & 181.mcf     & 63.4 & C  & 173.applu   & 47.9  \\
 2 & 186.crafty  & 27.3  & 2  & 177.mesa    & 11.2   \\
 0 & 197.parser  & 36.1  & 0  & 179.art     & 72.9 \\
 0 & 252.eon     & 8.7 & 0  & 183.equake  & 58.3  \\
 0 & 253.perlbmk & 24.0  & 0  & 187.facerec & 34.4  \\
 C & 254.gap     & 29.3  & C  & 188.ammp    & 46.4 \\
 I & 255.vortex  & 32.8  & F  & 189.lucas   & 41.7  \\
 N & 256.bzip2   & 36.7 & P  & 191.fma3d   & 33.8   \\
 T & 300.twolf   & 48.7 &    & 200.sixtrack& 5.7   \\  \cline{1-3}
 * & OpenOffice  & 10.5 &    & 301.apsi    & 44.4  \\  \cline{4-6}
   & RealPlayer  & 22.2  &    & 410.bwaves   & 42.98 \\ \cline{1-3}
 @ & SPECjbb2005 & 41.3 &    & 416.gamess   & 0.50 \\ \cline{1-3}
 S & 400.perlbench & 8.27  &   & 433.milc     & 96.54  \\
 P & 401.bzip2     & 33.60 & S & 434.zeusmp   & 34.14  \\
 E & 403.gcc       & 85.48 & P & 435.gromacs  & 4.92   \\
 C & 429.mcf       & 46.71 & E & 436.cactusADM& 24.36  \\
 2 & 445.gobmk     & 11.00  & C & 437.leslie3d & 84.84  \\
 0 & 456.hmmer     & 19.29  & 2 & 444.namd     & 1.19   \\
 0 & 458.sjeng     & 4.84  & 0 & 447.dealII   & 32.80  \\
 6 & 462.libquantum& 102.04 & 0 & 450.soplex   & 69.39 \\
 C & 464.h264ref   & 9.02  & 6 & 453.povray   & 0.10  \\
 I & 461.omnetpp   & 60.14 & C & 454.calculix & 3.67  \\
 N & 473.astar     & 31.16 & F & 459.GemsFDTD & 87.94  \\
 T & 483.xalancbmk & 29.16 & P & 465.tonto    & 26.06  \\ \cline{1-3}
 \multicolumn{3}{|l|}{* -- Desktop App}  & & 470.lbm      & 106.82  \\
 \multicolumn{3}{|l|}{@ -- Server  App}  & & 481.wrf      & 44.80   \\
 \multicolumn{3}{|l|}{ }                 & & 482.sphinx3  & 71.78  \\
 \hline
\end{tabular}
\vspace{-0.2cm}\end{table}

The overheads of the HMTT system include trace size, additional
memory references, execution time of I-Codes and kernel buffer for
collecting extra kernel data. Because applications usually generate
billions of traces during their execution periods, most trace sizes
are more than 10GB. The trace size is quite large, and large
capacity disks are demanded. Fortunately, it should not be a problem
because the disks are becoming larger and cheaper. On the other
hand, Figure \ref{fig:hmtt_design} illustrates that the HMTT system
adopt I-Codes to generate specific memory reference and collect
extra kernel data. In fact, there is almost no additional execution
time while I-Codes only generate specific memory references because
those specific memory references are less than one thousand of
normal memory references. For example, we have experimented on an
AMD machine and observed that applications' execution time is
increased by less than 1\% when I-Codes collect page table data upon
every page fault. To collect page table data, the assistant kernel
module requires to allocate a buffer which is less than 0.5\% of
total memory of traced system. Furthermore, these specific buffers
cannot induce significant influence because those references to the
specific buffers can be filtered.

\end{document}